\documentclass[manuscript,sn-basic,Numbered]{aastex631}

\setcitestyle{numbers, open={(}, close={)}}

\usepackage{hyperref}
\usepackage[T1]{fontenc}

\newcommand{\ms}{m\,s$^{-1}$}
\newcommand{\masy}{mas\,y$^{-1}$}
\newcommand{\mpl}{\mbox{$M_{\rm P}$}}
\newcommand{\rpl}{\mbox{$R_{\rm P}$}}
\newcommand{\rhopl}{\mbox{$\rho_{\rm P}$}}
\newcommand{\mstar}{\mbox{$M_{*}$}}
\newcommand{\rstar}{\mbox{$R_{*}$}}
\newcommand{\lstar}{\mbox{$L_{*}$}}

\newcommand{\mjup}{\mbox{$M_{\rm J}$}}
\newcommand{\rjup}{\mbox{$R_{\rm J}$}}

\newcommand{\mearth}{\mbox{$M_{\oplus}$}}
\newcommand{\rearth}{\mbox{$R_{\oplus}$}}
\newcommand{\msun}{\mbox{$M_{\odot}$}}
\newcommand{\rsun}{\mbox{$R_{\odot}$}}
\newcommand{\lsun}{\mbox{$L_{\odot}$}}
\newcommand{\gccc}{g\,cm$^{-3}$}

\newcommand{\teff}{\mbox{$T_{\rm eff}$}}
\newcommand{\feh}{\mbox{[Fe/H]}}

\newcommand{\logg}{$\log g$}
\newcommand{\tc}{$T_{\rm C}$}
\newcommand{\rprs}{\mbox{$R_{\rm P}/R_{*}$}}
\newcommand{\ars}{\mbox{$a/R_{*}$}}
\newcommand{\Nstar}{TOI-6894}
\newcommand{\Nstartic}{TIC-67512645}
\newcommand{\Nplanet}{TOI-6894\,b}
\newcommand{\NGAIAid}{391727828728624780}

\newcommand{\NgammaESP}{\mbox{$15826.4\pm6.1$}}
\newcommand{\NjitterESP}{\mbox{$0.1\pm7.1$}}
\newcommand{\NgammaSpirou}{\mbox{$15970\pm11$}}
\newcommand{\NjitterSpirou}{\mbox{$0.1\pm8.7$}}

\newcommand{\NRA}{\mbox{$11^{\mathrm h}33^{\mathrm m}52.5890{\mathrm s}$}} 
\newcommand{\NDec}{\mbox{$+12{\arcdeg}27{\arcmin}03.9373{\arcsec}$}} 
\newcommand{\Nplx}{\mbox{$13.684 \pm 0.053$}}
\newcommand{\Ndist}{\mbox{$72.96\pm0.29$}}

\newcommand{\NpropRA}{\mbox{$-146.897 \pm 0.056$}} 
\newcommand{\NpropDec}{\mbox{$22.227 \pm 0.053$}} 
\newcommand{\Nstarmass}{\mbox{$0.207 \pm 0.011$}} 
\newcommand{\Nstarradius}{\mbox{$0.2276\pm0.0057$}} 
\newcommand{\Nstardensity}{\mbox{$24.73\pm0.93$}} 
\newcommand{\Nstarlum}{\mbox{$0.00375 \pm 0.00033$}}
\newcommand{\Nteff}{\mbox{$3007 \pm 58$}} 
\newcommand{\Nmetal}{\mbox{$0.142 \pm 0.087$}} 
\newcommand{\Nlogg}{\mbox{$5.039 \pm 0.011$}} 

\newcommand{\NGAIAmag}{$16.2813\pm0.0011$}
\newcommand{\NGAIABpmag}{$18.125\pm0.018$}
\newcommand{\NGAIARpmag}{$14.9967\pm0.0022$}

\newcommand{\NJmag}{$13.169\pm0.023$}
\newcommand{\NHmag}{$12.486\pm0.022$}
\newcommand{\NKmag}{$12.207\pm0.021$}
\newcommand{\NWmag}{$12.020 \pm 0.023$} 
\newcommand{\NWWmag}{$11.842 \pm 0.022$}
\newcommand{\NWWWmag}{$11.16 \pm 0.15$}
\newcommand{\NTESSmag}{$14.9046 \pm 0.0078$}

\newcommand{\NteffODUSSEAS}{\mbox{$2960\pm66$}}
\newcommand{\NfehODUSSEAS}{\mbox{$-0.01\pm0.10$}}

\newcommand{\Nperiod}{\mbox{$3.37077196\pm0.00000059$}}
\newcommand{\Nperiodshort}{\mbox{$3.37$}}
\newcommand{\Nduration}{\mbox{$1.4220\pm0.0062$}}

\newcommand{\Ntc}{\mbox{$2460313.411670\pm0.000042$}}
\newcommand{\Necc}{\mbox{$0.029\pm0.030$}}%
\newcommand{\NeccUpperLim}{$0.094$} 
\newcommand{\Nmass}{\mbox{$0.168\pm0.022$}}
\newcommand{\Nmassearth}{\mbox{$53.4\pm7.1$}}
\newcommand{\Nmassratio}{\mbox{$(7.8\pm1.1) \times 10^{-4}$}}
\newcommand{\Nradius}{\mbox{$0.855\pm0.022$}}%
\newcommand{\Nradiusearth}{\mbox{$9.58\pm0.25$}}
\newcommand{\Ndensitycgs}{\mbox{$0.334\pm0.043$}} 
\newcommand{\Ngravp}{\mbox{$5.73\pm0.71$}} 
\newcommand{\Nrratio}{\mbox{$0.3860\pm0.0029$}}
\newcommand{\Nau}{\mbox{$0.02604\pm0.00045$}}%
\newcommand{\Naoverr}{\mbox{$24.59\pm0.31$}}%

\newcommand{\Nimpact}{\mbox{$0.177_{-0.040}^{+0.031}$}}%
\newcommand{\Ninc}{\mbox{$89.58_{-0.07}^{+0.10}$}}

\newcommand{\Nsemiamp}{\mbox{$65.5 \pm 8.3$}}
\newcommand{\NTeqA}{\mbox{$417.9 \pm 8.6$}} 
\newcommand{\Ninsol}{\mbox{$(7.54 \pm 0.60) \times 10^6$}}
\newcommand{\NinsolEarth}{\mbox{$5.50 \pm 0.44$}}

\newcommand{\NZp}{\mbox{$0.23 \pm 0.02$}}
\newcommand{\Nsolidmass}{\mbox{$12 \pm 2$}} 

\begin{document}

\title{A transiting giant planet in orbit around a 0.2-solar-mass host star}

\correspondingauthor{Edward M. Bryant}
\email{edward.m.bryant@warwick.ac.uk}

\author[0000-0001-7904-4441]{Edward M. Bryant}
\affiliation{Department of Space and Climate Physics, Mullard Space Science
Laboratory, University College London, Holmbury St Mary, RH5 6NT, UK}
\affiliation{Department of Physics, University of Warwick, Gibbet Hill Road, Coventry CV4 7AL, UK}

\author[0000-0002-5389-3944]{Andr\'es Jord\'an}
\affiliation{Facultad de Ingenier\'ia y Ciencias, Universidad Adolfo Ib\'a\~nez, Av. Diagonal las Torres 2640, Pe\~nalol\'en, Santiago, Chile}
\affiliation{Millennium Institute of Astrophysics (MAS), Nuncio Monseñor Sótero Sanz 100, Providencia, Santiago, Chile}
\affiliation{El Sauce Observatory, Obstech, Coquimbo, Chile}
 
\author[0000-0001-8732-6166]{Joel D. Hartman}
\affiliation{Department of Astrophysical Sciences, Princeton University, Princeton, NJ 08544, USA}

\author[0000-0001-6023-1335]{Daniel Bayliss}
\affiliation{Department of Physics, University of Warwick, Gibbet Hill Road, Coventry CV4 7AL, UK}
\affiliation{Centre for Exoplanets and Habitability, University of Warwick, Gibbet Hill Road, Coventry CV4 7AL, U}
 
\author[0000-0002-7444-5315]{Elyar Sedaghati}
\affiliation{European Southern Observatory (ESO), Av. Alonso de Córdova 3107, 763 0355 Vitacura, Santiago, Chile}

\author[0000-0003-1464-9276]{Khalid Barkaoui}
\affiliation{Astrobiology Research Unit, Universit\'e de Li\`ege, All\'ee du 6 Ao\^ut 19C, B-4000 Li\`ege, Belgium}
\affiliation{Department of Earth, Atmospheric and Planetary Science, Massachusetts Institute of Technology, 77 Massachusetts Avenue, Cambridge, MA 02139, USA}
\affiliation{Instituto de Astrof\'isica de Canarias (IAC), Calle V\'ia L\'actea s/n, 38200, La Laguna, Tenerife, Spain}

\author{Jamila Chouqar}
\affiliation{Astrobiology Research Unit, Universit\'e de Li\`ege, All\'ee du 6 Ao\^ut 19C, B-4000 Li\`ege, Belgium}
\affiliation{Oukaimeden Observatory, High Energy Physics and Astrophysics Laboratory, Cadi Ayyad University, Marrakech, Morocco}

\author[0000-0003-1572-7707]{Francisco J. Pozuelos}
\affiliation{Instituto de Astrof\'isica de Andaluc\'ia (IAA-CSIC), Glorieta de la Astronom\'ia s/n, 18008 Granada}

\author[0000-0002-5113-8558]{Daniel P. Thorngren}
\affiliation{Department of Physics \& Astronomy, Johns Hopkins University, Baltimore, MD 21210, USA}

\author[0009-0008-2214-5039]{Mathilde Timmermans}
\affiliation{Astrobiology Research Unit, Universit\'e de Li\`ege, All\'ee du 6 Ao\^ut 19C, B-4000 Li\`ege, Belgium}

\author[0000-0003-3208-9815]{Jose Manuel Almenara}
\affiliation{Univ. Grenoble Alpes, CNRS, IPAG, 38000 Grenoble, France}
\affiliation{Observatoire de Gen\`eve, Département d’Astronomie, Universit\'e de Gen\`eve, Chemin Pegasi 51b, 1290 Versoix, Switzerland}

\author[0000-0002-7924-3253]{Igor V. Chilingarian}
\affiliation{Center for Astrophysics \textbar \ Harvard \& Smithsonian, 60 Garden Street, Cambridge, MA 02138, USA}
\affiliation{Sternberg Astronomical Institute, M. V. Lomonosov Moscow State University, Universitetski prosp. 13, 119234 Moscow, Russia}

\author[0000-0001-6588-9574]{Karen A. Collins}
\affiliation{Center for Astrophysics \textbar \ Harvard \& Smithsonian, 60 Garden Street, Cambridge, MA 02138, USA}
 
\author[0000-0002-4503-9705]{Tianjun Gan}
\affiliation{Department of Astronomy, Tsinghua University, Beijing 100084, People's Republic of China}

\author[0000-0002-2532-2853]{Steve~B.~Howell}
\affiliation{NASA Ames Research Center, Moffett Field, CA 94035, USA}

\author[0000-0001-8511-2981]{Norio Narita}
\affiliation{Komaba Institute for Science, The University of Tokyo, 3-8-1 Komaba, Meguro, Tokyo 153-8902, Japan}
\affiliation{Astrobiology Center, 2-21-1 Osawa, Mitaka, Tokyo 181-8588, Japan}
\affiliation{Instituto de Astrof\'isica de Canarias (IAC), Calle V\'ia L\'actea s/n, 38200, La Laguna, Tenerife, Spain}
 
\author[0000-0003-0987-1593]{Enric Palle}
\affiliation{Instituto de Astrof\'isica de Canarias (IAC), Calle V\'ia L\'actea s/n, 38200, La Laguna, Tenerife, Spain}
\affiliation{Departamento de Astrof\'isica, Universidad de La Laguna (ULL), C/ Padre Herrera, 38206 La Laguna, Tenerife, Spain}

\author[0000-0002-3627-1676]{Benjamin V.\ Rackham}
\affiliation{Department of Earth, Atmospheric and Planetary Science, Massachusetts Institute of Technology, 77 Massachusetts Avenue, Cambridge, MA 02139, USA}
\affiliation{Kavli Institute for Astrophysics and Space Research, Massachusetts Institute of Technology, Cambridge, MA, USA}

\author[0000-0002-5510-8751]{Amaury H.M.J. Triaud}
\affiliation{School of Physics \& Astronomy, University of Birmingham, Edgbaston, Birmingham, B15 2TT, UK}

\author[0000-0001-7204-6727]{Gaspar \'A. Bakos}
\affiliation{Department of Astrophysical Sciences, Princeton University, Princeton, NJ 08544, USA}

\author[0000-0002-9158-7315]{Rafael Brahm}
\affiliation{Facultad de Ingenier\'ia y Ciencias, Universidad Adolfo Ib\'a\~nez, Av. Diagonal las Torres 2640, Pe\~nalol\'en, Santiago, Chile}
\affiliation{Millennium Institute of Astrophysics (MAS), Nuncio Monseñor Sótero Sanz 100, Providencia, Santiago, Chile}

\author[0000-0002-5945-7975]{Melissa J. Hobson}
\affiliation{Observatoire de Gen\`eve, Département d’Astronomie, Universit\'e de Gen\`eve, Chemin Pegasi 51b, 1290 Versoix, Switzerland}

\author[0000-0001-5542-8870]{Vincent Van Eylen}
\affiliation{Department of Space and Climate Physics, Mullard Space Science
Laboratory, University College London, Holmbury St Mary, RH5 6NT, UK}

\author[0000-0002-8388-6040]{Pedro J. Amado}
\affiliation{Instituto de Astrof\'isica de Andaluc\'ia (IAA-CSIC),
Glorieta de la Astronom\'ia s/n, 18008 Granada}

\author[0000-0002-0111-1234]{Luc Arnold}
\affiliation{Canada–France–Hawaii Telescope, 65-1238 Mamalahoa Hwy, Kamuela, HI 96743, USA}

\author[0000-0001-9003-8894]{Xavier Bonfils}
\affiliation{Univ. Grenoble Alpes, CNRS, IPAG, 38000 Grenoble, France}

\author[0000-0001-9892-2406]{Artem Burdanov}
\affiliation{Department of Earth, Atmospheric and Planetary Science, Massachusetts Institute of Technology, 77 Massachusetts Avenue, Cambridge, MA 02139, USA}
 
\author[0000-0001-9291-5555]{Charles Cadieux}
\affiliation{Universit\'e de Montr\'eal, D\'epartement de Physique, IREX, Montr\'eal, QC H3C 3J7, Canada}

\author[0000-0003-1963-9616]{Douglas A. Caldwell}
\affiliation{SETI Institute, Mountain View, CA 94043 USA}
\affiliation{NASA Ames Research Center, Moffett Field, CA 94035, USA}

\author[0000-0003-2036-8999]{Victor Casanova}
\affiliation{Instituto de Astrof\'isica de Andaluc\'ia (IAA-CSIC),
Glorieta de la Astronom\'ia s/n, 18008 Granada}

\author[0000-0002-9003-484X]{David Charbonneau}
\affiliation{Center for Astrophysics \textbar \ Harvard \& Smithsonian, 60 Garden Street, Cambridge, MA 02138, USA}

\author[0000-0002-2361-5812]{Catherine A. Clark}
\affiliation{NASA Exoplanet Science Institute, IPAC, California Institute of Technology, Pasadena, CA 91125, USA}

\author[0000-0003-2781-3207]{Kevin I.\ Collins}
\affiliation{George Mason University, 4400 University Drive, Fairfax, VA, 22030 USA}

\author[0000-0002-6939-9211]{Tansu Daylan}
\affiliation{Department of Physics, Washington University, St. Louis, MO 63130, USA}
\affiliation{McDonnell Center for the Space Sciences, Washington University, St. Louis, MO 63130, USA}

\author[0000-0002-3937-630X]{Georgina Dransfield}
\affiliation{School of Physics \& Astronomy, University of Birmingham, Edgbaston, Birmingham, B15 2TT, UK}

\author[0000-0002-9355-5165]{Brice-Olivier Demory}
\affiliation{Center for Space and Habitability, University of Bern, Gesellschaftsstrasse 6, CH-3012 Bern, Switzerland}

\author[0000-0002-7008-6888]{Elsa Ducrot}
\affiliation{LESIA, Observatoire de Paris, CNRS, Universit\'e Paris Diderot, Universit\'e Pierre et Marie Curie, 5 place Jules Janssen, 92190 Meudon, France}
\affiliation{AIM, CEA, CNRS, Universit\'e Paris-Saclay, Universit\'e de Paris, F-91191 Gif-sur-Yvette, France}

\author{Gareb Fernández-Rodríguez}
\affiliation{Instituto de Astrof\'isica de Canarias (IAC), Calle V\'ia L\'actea s/n, 38200, La Laguna, Tenerife, Spain}
\affiliation{Departamento de Astrof\'isica, Universidad de La Laguna (ULL), C/ Padre Herrera, 38206 La Laguna, Tenerife, Spain}
 
\author[0000-0002-9436-2891]{Izuru Fukuda}
\affiliation{Department of Multi-Disciplinary Sciences, Graduate School of Arts and Sciences, The University of Tokyo, 3-8-1 Komaba, Meguro, Tokyo, 153-8902, Japan}

\author[0000-0002-4909-5763]{Akihiko Fukui}
\affiliation{Komaba Institute for Science, The University of Tokyo, 3-8-1 Komaba, Meguro, Tokyo 153-8902, Japan}
\affiliation{Instituto de Astrof\'isica de Canarias (IAC), Calle V\'ia L\'actea s/n, 38200, La Laguna, Tenerife, Spain}
  
\author[0000-0003-1462-7739]{Micha\"el Gillon}
\affiliation{Astrobiology Research Unit, Universit\'e de Li\`ege, All\'ee du 6 ao\^ut 19, Li\`ege, 4000, Belgium}

\author[0009-0000-7274-7523]{Rebecca Gore}
\affiliation{Bay Area Environmental Research Institute, Moffett Field, CA 94035, USA}
\affiliation{NASA Ames Research Center, Moffett Field, CA 94035, USA}

\author[0000-0003-0030-332X]{Matthew J. Hooton}
\affiliation{Cavendish Laboratory, JJ Thomson Avenue, Cambridge CB3 0HE, UK}

\author[0000-0002-5978-057X]{Kai Ikuta}
\affiliation{Department of Multi-Disciplinary Sciences, Graduate School of Arts and Sciences, The University of Tokyo, 3-8-1 Komaba, Meguro, Tokyo, 153-8902, Japan}

\author[0000-0001-8923-488X]{Emmanuel Jehin}
\affiliation{Space Sciences, Technologies and Astrophysics Research (STAR) Institute, Universit\'e de Li\`ege, All\'ee du 6 Ao\^ut 19C, B-4000 Li\`ege, Belgium}

\author[0000-0002-4715-9460]{Jon M. Jenkins}
\affiliation{NASA Ames Research Center, Moffett Field, CA 94035, USA}

\author[0000-0001-8172-0453]{Alan M. Levine}
\affiliation{Department of Physics and Kavli Institute for Astrophysics and Space Research, Massachusetts Institute of Technology, Cambridge, MA 02139, USA}

\author{Colin Littlefield}
\affiliation{Bay Area Environmental Research Institute, Moffett Field, CA 94035, USA}
\affiliation{NASA Ames Research Center, Moffett Field, CA 94035, USA}

\author[0000-0001-9087-1245]{Felipe Murgas}
\affiliation{Instituto de Astrof\'isica de Canarias (IAC), Calle V\'ia L\'actea s/n, 38200, La Laguna, Tenerife, Spain}
\affiliation{Departamento de Astrof\'isica, Universidad de La Laguna (ULL), C/ Padre Herrera, 38206 La Laguna, Tenerife, Spain}

\author[0000-0002-5937-9655]{Kendra Nguyen}
\affiliation{Department of Astronomy, Yale University, New Haven, CT 06520, USA}

\author[0000-0001-5519-1391]{Hannu Parviainen }
\affiliation{Departamento de Astrof\'isica, Universidad de La Laguna (ULL), C/ Padre Herrera, 38206 La Laguna, Tenerife, Spain}
\affiliation{Instituto de Astrof\'isica de Canarias (IAC), Calle V\'ia L\'actea s/n, 38200, La Laguna, Tenerife, Spain}

\author[0000-0002-3012-0316]{Didier Queloz}
\affiliation{Cavendish Laboratory, JJ Thomson Avenue, Cambridge CB3 0HE, UK}
\affiliation{Department of Physics, ETH Zurich, Wolfgang-Pauli-Strasse 2, CH-8093 Zurich, Switzerland}

\author[0000-0002-6892-6948]{S.~Seager}
\affiliation{Department of Physics and Kavli Institute for Astrophysics and Space Research, Massachusetts Institute of Technology, Cambridge, MA 02139, USA}
\affiliation{Department of Earth, Atmospheric and Planetary Sciences, Massachusetts Institute of Technology, Cambridge, MA 02139, USA}
\affiliation{Department of Aeronautics and Astronautics, MIT, 77 Massachusetts Avenue, Cambridge, MA 02139, USA}

\author[0000-0002-2214-9258]{Daniel Sebastian}
\affiliation{School of Physics \& Astronomy, University of Birmingham, Edgbaston, Birmingham, B15 2TT, UK}

\author{Gregor Srdoc}
\affiliation{Kotizarovci Observatory, Sarsoni 90, 51216 Viskovo, Croatia}

\author[0000-0001-6763-6562]{R.~Vanderspek}
\affiliation{Department of Physics and Kavli Institute for Astrophysics and Space Research, Massachusetts Institute of Technology, Cambridge, MA 02139, USA}

\author[0000-0002-4265-047X]{Joshua N.\ Winn}
\affiliation{Department of Astrophysical Sciences, Princeton University, Princeton, NJ 08544, USA}

\author[0000-0003-2415-2191]{Julien de Wit}
\affiliation{Department of Earth, Atmospheric and Planetary Science, Massachusetts Institute of Technology, 77 Massachusetts Avenue, Cambridge, MA 02139, USA}
\affiliation{Kavli Institute for Astrophysics and Space Research, Massachusetts Institute of Technology, Cambridge, MA, USA}
 
\author[0000-0002-9350-830X]{Sebasti\'an Z\'u\~niga-Fern\'andez}
\affiliation{Astrobiology Research Unit, Universit\'e de Li\`ege, All\'ee du 6 ao\^ut 19, Li\`ege, 4000, Belgium}



\begin{abstract}

Planet formation models suggest that the formation of giant planets is significantly harder around low-mass stars, due to the scaling of protoplanetary disc masses with stellar mass. The discovery of giant planets orbiting such low-mass stars thus imposes strong constraints on giant planet formation processes. Here, we report the discovery of a transiting giant planet orbiting a \Nstarmass\,\msun\ star. The planet, \Nplanet, has a mass and radius of $\mpl = \Nmass\,\mjup \ (\Nmassearth\,\mearth)$  and $\rpl = \Nradius\,\rjup$, and likely includes  \Nsolidmass\,\mearth\ of metals. The discovery of \Nplanet\ highlights the need for a better understanding of giant planet formation mechanisms and the protoplanetary disc environments in which they occur. The extremely deep transits (17\% depth) make \Nplanet\ one of the most accessible exoplanetary giants for atmospheric characterisation observations, which will be key for fully interpreting the formation history of this remarkable system and for the study of atmospheric methane chemistry.
\end{abstract}


\section{Main Text}
\subsection{Introduction}

Core-accretion planet formation models predict that the ability to form a giant planet scales with the mass of the host star \cite{laughlin2004coreaccretion, burn2021ngppslowmassstars}. This is primarily a result of the fact these models suggest that a large amount of solid material in protoplanetary discs is necessary for the formation of giant planets, and observations have demonstrated that the mass of solid material in a protoplanetary disc scales with the mass of the star \cite{pascussi2016ppdisks, manara2023}. Therefore, it is expected that stars less massive than the Sun will form fewer giant planets \cite{burn2021ngppslowmassstars}. In fact, multiple studies have predicted that very low-mass stars ($\mstar \leq 0.3\msun$) will not be able to form giant planets \cite{idalin2005planetformation, liu2019pebbleaccretion, miguel2020lmstarplanetformation, mulders2021pebbleaccretion, burn2021ngppslowmassstars}. 

The discovery of exoplanets orbiting stars significantly less massive than the Sun\cite[e.g.][]{kanodia2023toi5205}, and determining their frequency of occurrence \cite[e.g.][]{gan2023earlymgiantplanetoccrates} is therefore a critical test of giant planet formation. Existing surveys have shown that giant planets must be very rare around mid-to-late M-dwarf stars \cite[e.g.][]{sabotta2021carmenes,pass2023}, but have not been able to provide robust occurrence rate measurements. 

To test the predictions of the formation theories we conducted a survey, using photometric data from the Transiting Exoplanet Survey Satellite (TESS; \cite{ricker2015tess}), to search for giant planets transiting low-mass host stars \cite{bryant2023lmstargiantplanetoccrates}. Amongst the planet candidates discovered by this survey was a candidate giant planet transiting the very low-mass star, \Nstar\ (note in \cite{bryant2023lmstargiantplanetoccrates} the candidate was listed by its TIC designation, TIC-67512645). 

\subsection{Results}
\subsubsection{Observations}
The $\Nstarmass\,\msun$ star \Nstar\ was initially observed by TESS from 2020 February 18 to March 18 in the Full Frame Images at a cadence of 30\,minutes. A candidate transiting planet signal at a period of \Nperiodshort\,days was reported by \cite{nguyen2022} and was subsequently independently identified by \cite{bryant2023lmstargiantplanetoccrates}. Further shorter cadence monitoring by TESS, at a 10-minute cadence from 2021 November 6 to December 30 and 2022 February 26 to March 26 and at a 2-minute cadence from 2023 November 11 to December 7, confirmed the presence of the transit signal and revealed it as a likely planet candidate (see Figure~1). Based on this additional monitoring and \cite{nguyen2022} the candidate was alerted as TOI-6894.01 by the TESS Science Office on 2024 February 1. 

Eclipsing binaries nearby to or in the background of the target star can blend into the photometric aperture and mimic a transiting exoplanet signal. The large pixel-scale of the TESS cameras means that there is a higher likelihood of this occurring, compared with other transit surveys. Additional transit observations of \Nplanet\ to investigate these scenarios were obtained using multiple ground-based telescopes (see Methods and Extended Data Figure 1). These observations revealed the transit signal is associated with the location of \Nstar, thereby ruling out nearby eclipsing binary scenarios. We analysed each transit individually and found that the depth of the transit does not significantly vary with the wavelength, thereby ruling out background eclipsing binary scenarios that lead to chromatic transits (see Methods and Extended Data Figure 2). These ground-based observations also improve the determination of the planet radius and orbital ephemeris and are included in the full analysis of the system.
Additionally, archival images dating back to 1952 show no background stellar contaminants at the current location of the system and high-angular resolution images show no associated sources in its immediate vicinity (see Methods and Extended Data Figure 3).  Photometric observations taken during the secondary eclipse reveal no deep eclipse signal (Extended Data Figure 4). All these together further validate the transit signal as genuine and likely due to a planetary companion.

We collected a mid-resolution near-IR spectrum of the host star using the FIRE spectrometer \cite{simcoe2013fire} on the Magellan telescope to assist with the stellar characterisation and provide a measure of the stellar metallicity (see Methods and Extended Data Figure 5).
High resolution spectroscopic observations obtained using the ESPRESSO spectrograph at the VLT \cite{pepe2021espresso} reveal a variation of the stellar radial velocity at an orbital period and phase consistent with the photometric transit signal (see Figure~1 and Extended Data Figure 6). Additional spectroscopic observations with the SPIRou spectrograph on the CFHT telescope \cite{Donati2020} corroborate this signal. We measure a radial velocity semi-amplitude of \Nsemiamp\,\ms\ which is consistent with a planetary nature for the transiting body. Combining this semi-amplitude with directly observable parameters from the transit light curves alone \cite{southworth2007loggp} we determine the transiting body's surface gravity to be $g_{\rm P} = \Ngravp$\,m\,s$^{-2}$, consistent with a planetary-mass object.

\subsubsection{Analysis}
We perform a joint analysis of all available observational data -- all the TESS and ground-based photometric data, the radial velocity measurements from ESPRESSO and SPIRou, broadband photometric measurements of the host star \Nstar\ and astrometric measurements from \textit{Gaia} \cite{GAIADR32021} -- to determine the stellar and planetary parameters. The combined ESPRESSO spectra and the FIRE spectrum were used to derive priors on the stellar atmospheric parameters and the analysis of the data was performed using a differential evolution Markov Chain Monte Carlo method (see Methods for details). The data and best fitting model are shown in Figure~1. (Full ground based photometry is shown in Extended Data Figure 1.) To quantitatively assess the likelihood of any blended eclipsing binary scenarios we model the available data as a blend between a bright M-dwarf star and a fainter blended eclipsing binary system (Methods). All blended binary scenarios provide significantly worse fits to the data than for the scenario of a single star with a transiting planet. As such, our analysis confidently confirms the nature of a the \Nstar\ system as a single star with a transiting planet, and we can confidently and quantitatively rule out all blended eclipsing binary scenarios.

From our joint analysis we find the host \Nstar\ to be a M$5.0\pm0.5$ dwarf star with a radius of \Nstarradius\,\rsun\ and a mass of \Nstarmass\,\msun, a very low mass to host a giant planet especially in the context of the known population of giant planets (see Figure~2). 
The low temperature of the host star (\teff$ = \Nteff K$) results in \Nplanet\ having a relatively cool equilibrium temperature of just \NTeqA\,K, assuming an albedo $A = 0.1$ and efficient heat redistribution. 
\Nplanet\ has a mass of \Nmass\,\mjup, which is just over half the mass of Saturn, and a radius of \Nradius\,\rjup, which is just larger than Saturn. Our analysis therefore reveals \Nplanet\ to be a low-density, giant planet. \Nplanet\ orbits its host star with a period of \Nperiod\,d. The analysis yields a measurement of the orbital eccentricity of \Necc\, with a 95\% confidence upper limit of \NeccUpperLim. The full set of derived parameters for both the planet and star are set out in Extended Data Tables 1 and 2. 

Using a retrieval framework for warm giant planets (see \cite{thorngren2019retrievals} and Methods) we model the interior structure of \Nplanet. We calculate a metal mass fraction -- the fraction of the total planet mass which is not hydrogen or helium -- of $Z_{\rm P} = \NZp$. From the measured stellar metallicity of $\feh = \Nmetal$ we calculate a stellar metal mass fraction of $Z_{\ast} = 0.0189 \pm 0.0037$, finding the planet to be metal-enriched compared to its host star, with a metal mass fraction a factor of twelve higher. We determine the metal mass content of \Nplanet\ to be $M_{\rm metal} = \Nsolidmass\,\mearth$. 

\subsection{Discussion}
\Nplanet\ joins an emerging population of giant planets with low-mass stars discovered through radial velocity observations -- LHS~3154~b \cite{stefansson2023lhs3154}, GJ~3512\,b \cite{morales2019gj3512}, GJ~3512\,c \cite{lopezsantiago2020}, and TZ~Ari~b \cite{quirrenbach2022} -- whose presence poses strong challenges to currently held formation theories. In particular the core-accretion model, one of the current leading mechanisms for giant planet formation, struggles to form planets with masses greater than 30\,\mearth\ around low-mass stars \cite{liu2019pebbleaccretion,miguel2020lmstarplanetformation,mulders2021pebbleaccretion,burn2021ngppslowmassstars}. The classic view of giant planet formation through core-accretion necessitates the formation of a massive core which then triggers a phase of runaway gas accretion \cite{pollack1996}. The primary hurdles to the formation of these planets are the limited amount of solid material within the protoplanetary disc with which to form a massive enough core, with lower-mass stars in general hosting lower-mass discs \cite{pascussi2016ppdisks}, along with the longer Keplerian timescales around these stars inhibiting the ability to form a massive enough core before the dispersal of the gas disc \cite{laughlin2004coreaccretion}.

With a sub-Saturn mass, however, \Nplanet\ may not have been required to undergo a phase of runaway gas accretion. Recent studies have proposed that sub-Saturn mass planets began their formation through a core-accretion process, but did not undergo runaway gas accretion \cite{helled2023}. Instead, an intermediate phase of heavy-element accretion occurred, accompanied by a steady accretion of gas onto the forming protoplanet \cite{helled2023}. Such a mechanism may provide a plausible pathway for the formation of \Nplanet\ without necessitating the rapid core formation or a runaway gas accretion phase.

Both the classic core-accretion and the sub-Saturn formation mechanisms would still require a suitable heavy element mass budget to be present in the protoplanetary disc to provide the $\Nsolidmass\,\mearth$ metal mass content of \Nplanet. The efficiency of giant planet formation, i.e., the fraction of the solid material in the disc available to be used to form the planet, has been estimated to be around 10\% \cite{lin2018}, following which the formation of \Nplanet\ would require a total of 120\,\mearth\, of solids to have been present in the disc.  From a sample of 70 Class II protoplanetary discs around stars within the mass range $0.15 - 0.25\,\msun$, the most massive has a dust mass of 58.6\,\mearth, and just a further four have a measured dust mass greater than the 12\,\mearth\ metal content of \Nplanet\ \cite{manara2023}. As such, from this simple mass budget argument it would initially appear that the formation of \Nplanet\ cannot be reconciled with the current sample of known protoplanetary discs. 

However, there are a number of important caveats to this argument.  First, these disc masses are calculated from the emission flux received from discs at millimetre wavelengths. Solid material in the disc in the form of centimetre-sized or larger pebbles would be undetectable through these observations, leading to an underestimation of the disc dust mass \cite{liu2022dustmassunderestimation}. Similarly, observations of younger Class 0 and I discs have also shown these discs to have dust masses an order of magnitude higher than Class II discs \cite{tychoniec2020}, and it has been theorised that large protoplanets may form during the Class 0/I phase of the protoplanetary disc \cite{nixon2018earlyplanetformation}. Furthermore, the current estimates of the formation efficiency are uncertain and depend on a number of poorly constrained characteristics of the protoplanetary disc \cite{chachanlee2023}. Moreover, given the rarity of planets such as \Nplanet\ \cite{bryant2023lmstargiantplanetoccrates}, and given the small sample size of low-mass star discs studied, it is not unexpected that we are yet to discover a massive enough disc to easily explain the formation of \Nplanet. Therefore, it is plausible that \Nplanet\ could have formed through a core-accretion-like mechanism, either the classic picture or the sub-Saturn variation. Further understanding of these formation mechanisms and the nature of protoplanetary discs around these low-mass stars is required to fully reconcile this planet with the formation theory. \Nplanet\ will stand as a key benchmark planet for anchoring future theoretical studies in these areas.

An alternative pathway for the formation of massive planets is direct formation through condensation from a gravitationally unstable disc \cite{boss1997diskinstability}. This mechanism has been shown to be capable of forming massive planets around low-mass stars, including the planet GJ~3512~b \cite{morales2019gj3512}. However, simulations provide differing conclusions on the feasibility of forming a planet like \Nplanet. One set of simulations of planet formation around low-mass stars produced very massive planets with masses $\geq 2\,\mjup$ \cite{mercerstamatellos2020diskinstability}. These simulations would therefore suggest that \Nplanet\ could not have been formed through this mechanism. Conversely, a different suite of simulations demonstrated this mechanism to be capable of forming exoplanets with masses in the range $0.1 - 0.3\,\mjup$ around 0.2\,\msun\ protostars \cite{bosskanodia2023mdwarfDI}. These simulations would therefore suggest this mechanism as a plausible formation pathway for \Nplanet. As the authors of the second study note, these two suites of simulations had large differences in the initial conditions they assumed for the protoplanetary discs. Therefore, this mechanism remains a plausible formation pathway for \Nplanet, although further understanding of the nature of protoplanetary discs will be required to fully interpret the formation of \Nplanet\ through this mechanism.

One potential hurdle to explaining the formation of \Nplanet\ through gravitational instability comes from recent planet synthesis simulations \cite{forgan2018}, which did not form any planet with a core mass greater than $5\,\mearth$. This is significantly less than the $12\pm2\,\mearth$ metal mass content of \Nplanet. However, we note that these simulations did not consider the subsequent accretion of solids onto the formed fragments, and so these simulations underestimate the final metal mass content of the planets. There is also the possibility that a substantial fraction of the metal constituents of \Nplanet\ may be present in its atmosphere and may have been delivered through the capture of planetesimals by the protoplanet \cite{helled2006}. Such a dispersal of the metal content within \Nplanet\ would reconcile the nature of the planet with potential formation through gravitational instability. Atmospheric characterisation through transmission spectroscopy will enable us to measure the atmospheric metallicity of \Nplanet\ \cite{blootmiguel2023}, thereby also providing a more robust measurement of the core mass, whose estimation from interior structure models based off mass and radius alone is degenerate with the atmospheric metallicity \cite{blootmiguel2023}. Atmospheric characterisation can therefore provide a pathway for determining whether gravitational instability remains a plausible formation mechanism for \Nplanet.

\Nplanet\ is a key exoplanet for further exo-atmospheric investigations, beyond untangling the puzzling question of its formation. The planet's equilibrium temperature makes it an intermediate object between the hot Jupiters that are being extensively observed by ground-based and space-based facilities \cite{Sedaghati2017,Rustamkulov2023}, and the cold gas-giants of our own solar system, Jupiter and Saturn. Based on its stellar irradiation, we expect that the planet's atmosphere is dominated by methane chemistry \cite{Zahnle2014,Fortney2020}. This alone would already make \Nplanet\ a very valuable new discovery since few such examples have been published \cite{bell2023}, but what makes it truly special compared to previously studied objects such as WASP-80\,b \cite{triaud2015,bell2023} is the combination of its particularly small host star, short orbital period, and low planetary density for its cool equilibrium temperature. Combined, these make \Nplanet\ an extremely accessible giant planet with a low-mass host star for transmission spectroscopy observations (Figure~3). Considering the transmission spectroscopy metric \cite[TSM;][]{kempton2018tsm}, a measure of the predicted signal-to-noise achieved for transmission spectroscopy observations, \Nplanet\ has a TSM value of $356\pm58$ which is the highest of any giant planet with an equilibrium temperature $T_{\rm eq} \leq 900 K$ or a host star mass $\mstar \leq 0.7 \msun$ (see Figure 3, Extended Data Figure 7, and Methods). Atmospheric models with and without clouds reveal that spectroscopic features in the transmission and emission spectra have expected amplitudes in excess of many planets' primary transits (Methods and Extended Data Figure 8). The detection of spectral features, the determination of the presence of clouds, and the measurement of the atmospheric metallicity are possible even with medium-sized ground-based telescopes or from just a single transit observation with JWST (Methods). \Nplanet\ will therefore be a benchmark exoplanet for the study of methane-dominated atmospheres.

As a very low-mass star hosting a transiting giant planet, the \Nstar\ system stands as a benchmark system for the understanding of giant planet formation, and challenging the current theories which struggle to explain its presence. The system is also highly amenable to transmission spectroscopy observations, through which we will be able to precisely determine both the atmospheric and interior composition of \Nplanet. The \Nstar\ system therefore stands to be a key exoplanetary system for determining the formation histories of giant planets, especially those with the lowest-mass host stars.

\begin{figure}
    \centering
    \includegraphics[width=\linewidth]{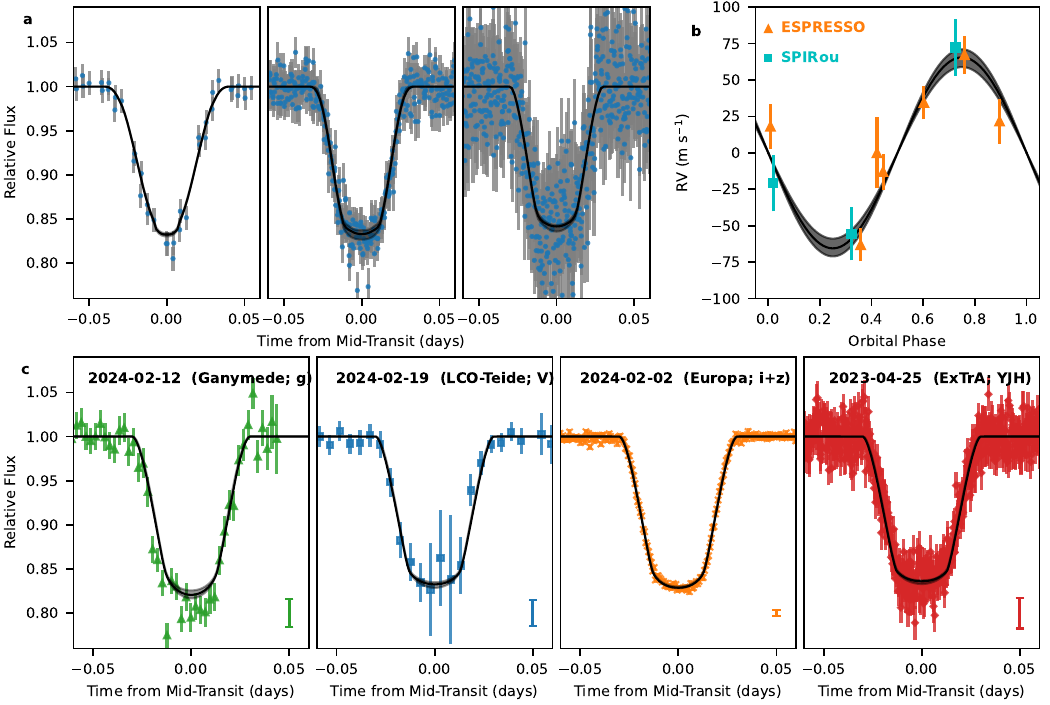}
    \caption{Observational data and best-fitting models (black line in all panels). \textbf{a.} Phase-folded TESS photometric data at a cadence of 30-min (left), 10-min (middle), and 2-min (right) (blue points). \textbf{b.} Phase-folded RV data from ESPRESSO (orange triangles) and SPIRou (cyan squares). \textbf{c.} Selected ground-based follow-up photometric data. The panel annotations give the night on which the observations were taken, the facility that performed the observations where Europa and Ganymede are two SPECULOOS-South nodes, and the observing filter used. For all plots, the errorbars plotted are the reported uncertainties for each data point and the grey shaded regions give the $1\sigma$ uncertainty on the model. The errorbars in the bottom-right corners of the lower four panels denote the median errorbar for the plotted observation. Note that all follow-up photometry, included the observations not plotted here, was included in the analysis; all follow-up photometry is plotted in Extended Data Figure 1.}
    \label{fig:main1}
\end{figure}

\begin{figure}
    \centering
    \includegraphics[width=\linewidth]{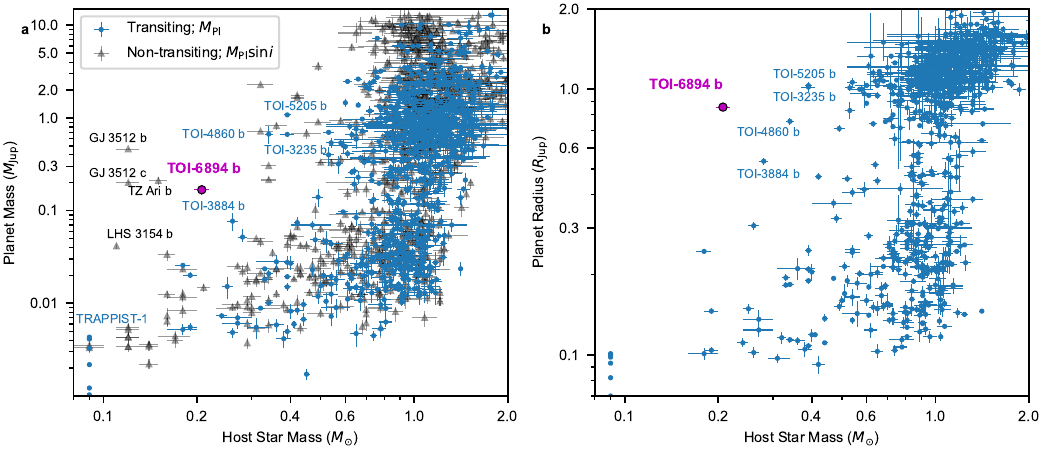}
    \caption{Placing \Nplanet\ in the context of known transiting planets. \textbf{a} Masses or minimum masses of the known population of planets discovered through the transit or radial velocity method as a function of mass of the host star (data taken from the NASA Exoplanet Archive, accessed 16 May 2024). We plot transiting planets for which we have an absolute mass measurement as the blue circles and the non-transiting RV planets for which we have just a lower limit on the mass as the gray triangles. \Nplanet\ is plotted as the purple circle. The planets mentioned in the text and the transiting giant planets around mid M-dwarf stars are labelled for reference. The errorbars are the one sigma uncertainty ranges on the plotted parameters.  \textbf{b} The same sample but now showing the planet radii. }
    \label{fig:main2}
\end{figure}

\begin{figure}
    \centering
    \includegraphics[width=\linewidth]{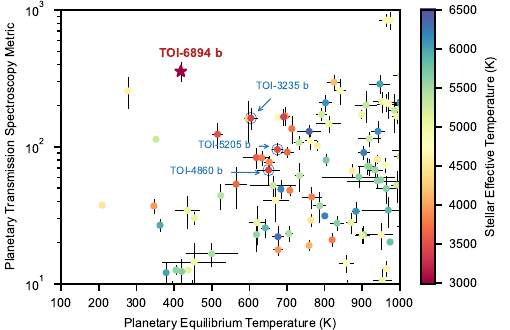}
    \caption{Atmospheric characterisation potential of \Nplanet. We plot the Transmission Spectroscopy Metric \cite[TSM; see ][ for more details]{kempton2018tsm} -- a metric which provides an estimate of the expected signal-to-noise for transmission spectroscopy observations -- of known giant planets as a function of the planetary equilibrium temperature. The colour of the points denotes the effective temperature of the host star and the star symbol denotes \Nplanet. The values and error bars provided for the known population are calculated from the values provided in the NASA Exoplanet Archive. We highlight three known transiting giant planets with mid-M-dwarf host stars with the blue circles and arrows.}
    \label{fig:main3}
\end{figure}

\begin{table}
    \centering
    \caption{Stellar Properties for \Nstar}
    {\renewcommand{\arraystretch}{0.8}
\begin{tabular}{lcc} 
	\hline
	\multicolumn{3}{l}{\textbf{Identifiers}}\\
	\Nstar \\
	\Nstartic \\
	\textit{Gaia} DR2\,\NGAIAid \\
    \hline
    \textbf{Property}	&	\textbf{Value}		&\textbf{Source}\\
    \hline
    \multicolumn{3}{l}{\textbf{Astrometric Properties}}\\
    R.A.		&	\NRA			&{\em Gaia}	DR3\\
	Dec.		&	\NDec			&{\em Gaia}	DR3\\
    $\mu_{{\rm R.A.}}$ (\masy) & \NpropRA & {\em Gaia} DR3\\
	$\mu_{{\rm Dec.}}$ (\masy) & \NpropDec & {\em Gaia} DR3\\
	Parallax (mas)   &   \Nplx           &{\em Gaia} DR3\\
    \\
    \multicolumn{3}{l}{\textbf{Photometric Properties}}\\
	TESS (mag)  &\NTESSmag     &TIC8\\
    \textit{Gaia} G (mag)&\NGAIAmag	&{\em Gaia} DR3\\
    \textit{Gaia} B$_P$ (mag)  &\NGAIABpmag    &{\em Gaia} DR3\\
    \textit{Gaia} R$_P$ (mag)  &\NGAIARpmag    &{\em Gaia} DR3\\
    J (mag)		&\NJmag		&2MASS	\\
   	H (mag)		&\NHmag		&2MASS	\\
	K (mag)		&\NKmag		&2MASS	\\
    W1 (mag)	&\NWmag		&WISE	\\
    W2 (mag)	&\NWWmag	&WISE	\\
    W3 (mag)	&\NWWWmag	&WISE	\\
    \\
    \multicolumn{3}{l}{\textbf{Derived Properties}}\\
    \teff\ (K)    & \Nteff              &This work (Methods)\\
    \feh\	     & \Nmetal			    &This work (Methods)\\
    \logg               & \Nlogg			&This work (Methods)\\
    \mstar (\msun) & \Nstarmass	        &This work (Methods)\\
    \rstar (\rsun) & \Nstarradius	            &This work (Methods)\\
    $\rho_*$ (\gccc) & \Nstardensity              & This work (Methods)\\
    \lstar (\lsun) & \Nstarlum              &This work (Methods)\\
    Distance (pc)	&  \Ndist	                &This work (Methods)\\
	\hline
    \multicolumn{3}{l}{2MASS \cite{2MASS}; {\em Gaia} DR3 \cite{GAIADR32021}; TIC8 \cite{stassun2019tic8}; WISE \cite{WISE}}
\end{tabular}}
\label{tab:stellar}
\end{table}

\begin{table}
    \centering
    \caption{Planetary Properties for \Nplanet}
    \begin{tabular}{lccc}
\toprule
    Name & Symbol & Unit & Value \\
    \noalign{\smallskip} 
    \hline
    \noalign{\smallskip} 
    Transit Mid-Point Time & \tc & BJD (TDB) & \Ntc \\
    \noalign{\smallskip} 
    Orbital Period & $P$ & days & \Nperiod \\
    \noalign{\smallskip} 
    Radius Ratio & \rprs & & \Nrratio \\
    \noalign{\smallskip} 
    Scaled Semi-major Axis & $a / R_*$ & & \Naoverr \\
    \noalign{\smallskip} 
    Impact Parameter & $b$ & & \Nimpact \\
    \noalign{\smallskip} 
    Orbital Inclination & $i$ & degrees & \Ninc \\
    \noalign{\smallskip} 
    Transit Duration & $T_{\rm dur}$ & hours & \Nduration \\
    \noalign{\smallskip} 
    RV Semi-Amplitude & $K$ & \ms & \Nsemiamp \\
    \noalign{\smallskip} 
    Orbital Eccentricity & $e$ & & \Necc\ ($\leq \NeccUpperLim$)\\
    \noalign{\smallskip} 
    Planet Radius & \rpl & \rjup & \Nradius \\
    \noalign{\smallskip} 
    Planet Mass & \mpl & \mjup & \Nmass \\
    \noalign{\smallskip} 
    Planet Bulk Density & \rhopl & \gccc & \Ndensitycgs \\
    \noalign{\smallskip}
    Planet-to-Star Mass Ratio & \mpl / \mstar & & \Nmassratio \\
    \noalign{\smallskip} 
    Planet Surface Gravity & $g_{\rm P}$ & m\,s$^{-1}$ & \Ngravp \\
    Semi-major Axis & $a$ & AU & \Nau\\
    \noalign{\smallskip} 
    Planet Irradiation Flux & $S$ & erg\,cm$^{-2}$\,s$^{-1}$ & \Ninsol \\
    \noalign{\smallskip} 
    Planet Equilibrium Temperature$^{a}$ & $T_{\rm eq}$ & K & \NTeqA \\
    \noalign{\smallskip} 
    Transmission Spectroscopy Metric & TSM & & $356 \pm 58$ \\
    \noalign{\smallskip}
    \toprule
    \multicolumn{4}{l}{a - assuming albedo $A = 0.1$}
\end{tabular}
\label{tab:planet}
\end{table}

\section{Methods}
\textbf{TESS Observations}\\
\Nstar\ (\Nstartic) was observed by the Transiting Exoplanet Survey Satellite \cite[TESS; ][]{ricker2015tess} during both the primary and extended mission. In the primary mission, \Nstar\ was observed in sector~22 (18 February to 18 March 2020) and in the extended mission \Nstar\ was observed in sectors 45, 46, and 49 (06 November to 30 December 2021 and 26 February to 26 March 2022). Across all sectors, \Nstar\ was observed in the Full-Frame-Images, and so the TESS photometry is available at a cadence of 30\,minutes for sector~22 and 10\,minutes for the extended mission sectors. The TESS FFI photometry was processed by the TESS Science Processing Operation Center \cite[SPOC; ][]{jenkins2016spoc} and we access the data through the TESS-SPOC High Level Science Product  \cite{caldwell2020tess_spoc}. For our analysis, we used the \textsc{PDCSAP} light curves, which have been processed to remove spacecraft related instrumental systematics \cite{smith2012pdc,stumpe2012pdc,stumpe2014pdc}; the TESS light curves for \Nstar\ are displayed in Figure~1. We display a cutout pixel image of the area surrounding \Nstar\ in Supplementary Figure 1. \\

\textbf{TESS candidate detection}\\
\Nstar\ was included in a systematic transit search for giant planets with low-mass host stars in the TESS primary mission FFI data \cite{bryant2023lmstargiantplanetoccrates}. In short, this search detected periodic transit-like signals using the \textsc{astropy} implementation of the Box-fitting Least Squares algorithm \cite{kovacs2002bls,astropy3_2022}, excluded clear false-positive scenarios, and performed a transit fitting analysis to identify likely giant planet candidates. Following these automated steps and some further manual vetting \Nplanet\ was identified as a good quality giant planet candidate \cite{bryant2023lmstargiantplanetoccrates}. The TESS SPOC independently identified the signature of TOI-6894b in transit searches of the FFI data from sectors 45, 46, and 49 using an adaptive matched-filter \cite{jenkins2002, jenkins2010, jenkins2020}, and after vetting the results of sector 49 with a modified version of TESS-ExoClass (TEC; \url{https://github.com/christopherburke/TESS-ExoClass}) for FFI targets \cite{caldwell2020tess_spoc} reported TOI-6894b as a candidate \cite{nguyen2022}. The difference image centroid analysis \cite{Twicken2018} for sector 46 constrained the location of the target star to be within $4.3\pm2.5$ arcsec of the transit source, significantly reducing the possibility of a nearby blended eclipsing binary scenario. \Nplanet\ was made a TESS Object of Interest (TOI) on 1st February 2024. \\

\textbf{ExTrA Observations}\\
A full transit of \Nplanet\ was observed by ExTrA \cite{bonfils2015extra}, a low-resolution near-infrared (0.85 -- 1.55 $\mu$m) multi-object spectrograph, on 2023 April 25. ExTrA is fed by three 60\,cm diameter telescopes and is located at ESO's La Silla Observatory in Chile. 
Five fibers are positioned in the focal plane of each telescope to select light from the target and four comparison stars. Due to the faintness of the target ($J = 13.2$\,mag), we utilised the low-resolution mode of the spectrograph ($R$$\sim$20) and employed the fibers with a 4\arcsec\ aperture to minimise the contribution of sky emission. The resulting ExTrA data were analyzed using custom data reduction software.
The transit light curves from the three ExTrA telescopes are presented in Figure~1 and Extended Data Figure 1. \\

\textbf{SPECULOOS Observations}\\
Six full transits of \Nplanet\ were observed using various telescopes in the SPECULOOS \cite{Jehin2018Msngr,Delrez2018,Sebastian_2021AA} 1m0-network located at ESO Paranal Observatory in Chile and Teide Observatory in Tenerife \cite{Burdanov_SPECULOOS_North_2022PASP}. All telescopes are equipped with a deep-depletion Andor iKon-L $2k \times 2k$ CCD camera with a pixel scale of $0.35$\arcsec, resulting in a total field of view of $12\arcsec \times 12\arcsec$.
We collected the data during transits of \Nplanet\ on the nights of 2024 February 02, 12, 19 in the $I+z'$, Sloan-$g'$, Sloan-$r'$ and Sloan-$z'$ filters, and data during an occultation of \Nplanet\ on the night of 2024 February 07 in the Sloan-$z'$ filter. 
Science image processing and photometric extraction were performed using the \texttt{PROSE} pipeline (\cite{2022_prose}; \url{https://github.com/lgrcia/prose}). The SPECULOOS data are detrended using external systematics variations related to time, the FWHM of the PSF, the sky background, the airmass and the X and Y pixel position. All the SPECULOOS transit photometry is plotted in Extended Data Figure 1 and a selection is plotted in Figure~1. The SPECULOOS occultation observation is plotted in Extended Data Figure 4. \\

\textbf{TRAPPIST Observations}\\
A full transit of \Nplanet\ was observed with the TRAPPIST-South \cite{Jehin2011,Gillon2011} telescope on 2024 February 12 in the Blue-Blocking ($BB$) filter with an exposure time of 140s. It is a 60-cm robotic Ritchey-Chretien telescope installed at ESO's La Silla Observatory in Chile. It is equipped with a thermoelectrically cooled 2K$\times$2K FLI Proline CCD camera with a pixel-scale of 0.65\arcsec and a field of view of $22\arcmin\times22\arcmin$ \cite{Jehin2011,Gillon2011}. Science images processing and photometric measurements were performed using the {\tt PROSE} pipeline. The TRAPPIST photometry is plotted in Extended Data Figure 1. \\

\textbf{Sierra Nevada Observatory (OSN/T150) Observations}\\
We observed TOI-6894\,b on 2024 February 19 using the T150 at the Sierra Nevada Observatory in Granada (Spain). The T150 is a 150-cm Ritchey-Chr\'etien telescope equipped with a thermoelectrically cooled 2K$\times$2K Andor iKon-L BEX2DD CCD camera with a field of view of $7.9'\times7.9'$ and pixel scale of 0.232". We used the Johnson-Cousin $I$ and $V$ filters simultaneously with exposure times of 120 and 90\,s, respectively. The photometric data were extracted using the {\tt AstroImageJ} package \cite{Collins2017} and are plotted in Extended Data Figure 1.\\

\textbf{LCOGT Observations}\\
\Nstar\ was also observed from the South African (SAAO) and Tenerife (Teide) nodes of the Las Cumbres Observatory Global Telescope network \cite[LCOGT;][]{Brown2013} using the 1-m telescopes on 2024 February 19. Both observations were carried out alternately in $V$ and $z_s$ band with exposure times of 300s and 70s, covering the full transits. The observations were done with the Sinistro cameras, which have a field of view of $26'\times26'$ and a pixel scale of $0.389\arcsec$. The raw images were automatically calibrated using the \textit{BANZAI} pipeline \cite{McCully2018}. We then performed the photometric analysis using the \textit{AstroImageJ} software \cite{Collins2017} with a 8-pixel ($3.1\arcsec$) and 5-pixel ($1.9\arcsec$) aperture. The estimated PSF of the two observations are 1.85$\arcsec$ and 1.65$\arcsec$, respectively. All the LCO photometry is plotted in Extended Data Figure 1 and the $V$ band photometry is also plotted in Figure~1. \\

\textbf{MUSCAT2 Observations}\\
A full-transit observation of \Nplanet\ was collected on UT 2024 February 19 using MuSCAT2 \cite{Narita2019} mounted on the 1.52 m Telescopio Carlos S\'{a}nchez at Teide Observatory, Tenerife, Spain. MuSCAT2 is a multicolour imager with a field of view of $7.4'\times7.4'$ and a pixel scale of $0.44\arcsec$. The observation was carried out simultaneously in four bands ($g$, $r$, $i$, and $z_s$). However, the $g$ and $r$-band data have a low signal-to-noise ratio due to the large scatter induced by the presence of  clouds. Therefore, we excluded these two data sets in our analysis. The $i$ and $z_s$-band data is also impacted by the clouds but are still maintain a sufficient signal-to-noise ratio to be usefully included in the analysis. We carried out aperture photometry using the MuSCAT2 pipeline \cite{Parviainen2020} after dark-frame and flat-field calibration. The pipeline automatically finds the optimised aperture to minimise the photometric dispersion, and then fits a transit model after accounting for the instrumental systematic effects. The MuSCAT2 data are plotted in Extended Data Figure 1. \\

\textbf{Magellan/FIRE}\\
\Nstar\ was observed on the night of 2024 February 26 with the Folded-Port InfRared Echellette (FIRE) \cite{simcoe2013fire} intermediate-resolution spectrograph operated at the 6.5-m Magellan Baade telescope, Las Campanas Observatory, Chile. We utilised a 0.6$\times$7~arcsec slit that provided a spectral resolving power $R=4500$ in the wavelength range $0.82<\lambda<2.5 \mu$m. We collected four 5~min-long exposures (a total integration of 20~min on source) with a $\pm1.5$~arcsec nodding along the slit in the `ABAB' pattern under 0.65~arcsec FWHM $J$-band atmospheric image quality. A telluric standard star 69~Leo (A0V) was observed right before the target and was used for flux calibration purposes. We reduced the FIRE spectra of \Nstar\ using the FIRE bright source pipeline \cite{2020ASPC..522..623C} that outputs a flux calibrated telluric-corrected spectrum merged from all 21 available Echelle orders. A telluric correction algorithm \cite{2023ApJS..266...11B} fits an observed stellar spectrum against a non-negative linear combination of synthetic stellar templates and a grid of earth atmospheric transmission models computed using {\sc ESO SkyCalc} \cite{2014A&A...567A..25N} for ESO La Silla, an observing site with very similar properties located geographically nearby to Las Campanas. The algorithm adjusts the final wavelength solution using telluric absorption lines to the final precision of about $\approx 0.3$~km~s$^{-1}$. A spectrum is presented in Extended Data Figure 5. \\

\textbf{ESPRESSO Observations}\\
We obtained spectroscopic observations of \Nstar\ using the ESPRESSO \cite{pepe2021espresso} high resolution, fiber-fed, cross-dispersed, echelle spectrograph in order to monitor the radial velocity (RV) variations due to the orbit of its companion and measure the mass of this transiting companion, thereby confirming its planetary nature. ESPRESSO is mounted at the Incoherent Combined Coudé Facility (ICCF) of ESO's Very Large Telescope (VLT), Paranal observatory in Chile. The observations were performed in the High Resolution (HR; $\mathcal{R}\sim140\,000$) mode, as part of a program dedicated to measuring the masses of giant planets around low-mass host stars (108.22B4.001; PI Jordan). We obtained seven spectra of \Nstar\ between 2022 February 3 and 8, using an exposure time of 2400s for each observation. The ESPRESSO DRS pipeline \cite[v2.3.5;][]{sosnowska2015espressoDRS,modigliani2020espressoDRS}, as implemented within the EsoReflex environment \cite{freudling2013esoreflex}, was used to reduce the spectra. The RVs were measured using ESPRESSO's dedicated Data Analysis Software (DAS; v1.3.6). It measures the RVs by fitting a Gaussian model to the cross correlation function (CCF). The CCF is derived by the DAS using an M4 stellar template which most closely matches the spectral type of the host star. The ESPRESSO RVs are listed in Extended Data Table 2 and presented in Figure~1 and Extended Data Figure 6. In addition to this approach, we also measured the RVs using the SERVAL pipeline \cite{SERVAL}, which employs the template-matching technique to obtain stellar RVs, which are consistent with the previous method within the uncertainties. We observed no significant correlation of the RV residuals with any of the activity indicators measured either by the DAS or SERVAL, which include the bisector span, CaII\,$\log{R^\prime_{HK}}$, the differential line width or the chromatic index. We also compute the $H_\alpha$ index at both 0.6 and 1.6\,\AA~using the ACTIN2 toolkit \cite{ACTIN2018, ACTIN2021}, again finding no significant. We do however note relatively large variations in the absolute values of the bisector span, which are due to noise in the CCF, as well as the complex shape of those functions. We computed the periodogram of the ESPRESSO RVs. We found that there was a signal at the planet orbital period, although we note that the significance from the RVs alone of this signal is low. So while this is a significant detection given our prior knowledge of the planet's period from the TESS photometry, in the case of a blind RV search we note that more RVs would be required to achieve a confident blind detection. \\

\textbf{SPIRou Observations}\\
We obtained three spectroscopic observations between 2024 February 22 and 24 (Program 24AD02; PI Gan) for \Nstar\ using SPIRou \cite[SpectroPolarim\`etre InfraROUge;][]{Donati2020}, installed on the 3.6m Canada-France-Hawaii Telescope (CFHT). SPIRou is a fiber-fed near-infrared high resolution ($\mathcal{R}\sim75\,000$) spectropolarimeter with a wavelength coverage between 0.98 and 2.5\,$\mu$m. Since the host star is faint in the H band, we chose to conduct the observations in the Dark mode without simultaneous drift calibration with the thermalised Fabry-P\'erot (FP) etalon, in order to avoid contamination. All observations were collected with an exposure time of 1800 seconds under an environment of airmass around 1.0 and seeing about $0.6\arcsec$, achieving S/N values of 89, 87, and 84 at order 44 (2.16 to 2.22 $\mu m$).

We reduced the data using {\tt APERO} \cite{Cook2022} and extracted RV values through the line-by-line (LBL) method from the telluric corrected spectra \cite{Artigau2022}. The final RVs are the error-weighted average of all valid per-line velocities. The LBL method has been used in several recent TESS-related works to determine the mass of planets (e.g., TOI-1759\,b, \cite{Martioli2022}; TOI-2136\,b, \cite{Gan2022}; TOI-1452\,b, \cite{Cadieux2022}; TOI-1695\,b, \cite{Kiefer2023}; TOI-4201\,b, \cite{Gan2023}). The SPIRou RVs are plotted in Figure~1 and Extended Data Figure 6 and listed in Extended Data Table 2. It must be noted that there exists a systematic offset between the systemic velocity values obtained from the ESPRESSO and the SPIRou observations (Extended Data Table 2), which is due to the differences between the instrumental zeropoints, as well as the wavelength coverage of the two instruments. \\

\textbf{Archival Imaging}\\
Due to the high proper motion of \Nstar\ (148.6\,\masy), archival imaging provides a useful check on line-of-sight blended neighbours. The 48-Inch Oschin Telescope at Palomar Mountain, California imaged \Nstar\ on the night of 1952 January 31 as part of the Palomar Observatory Sky Survey.  The image was a 1-hour exposure using the R-band filter.  We accessed the digitised plate via the Space Telescope Science Institute's Digitized Sky Survey (\url{https://archive.stsci.edu/dss}).  The image shows \Nstar\ approximately 10\arcsec\ to the east of its current location (see Extended Data Figure 3), in agreement with the star's proper motion as measured by Gaia (see Table~\ref{tab:stellar}).  Analysis of the Palomar image shows no background source at the present position of \Nstar\ to the sensitivity of the photographic plate, which we estimate by cross-matching Gaia DR3 point sources to 5-sigma detections on the image.  We can therefore rule out blended background sources to a magnitude limit of $G = 19.5$ mag. \\

\textbf{High Contrast Imaging}\\
While blended background sources are ruled out from archival imaging, there is still the possibility of blending due to a co-moving companion. To investigate the possible multiplicity of \Nstar, we obtained high-resolution imaging using the `Alopeke speckle imager \cite{scott2021} at Gemini North on 2024 May 22 (UT). We used 562/44 nm and 832/40 nm for the blue and red cameras, respectively. In each channel, we obtained 17,000 individual 60 ms frames, for a total integration time of 17 min in each band. Immediately thereafter, we observed a nearby star at a similar airmass in order to measure the speckle-transfer function. The data were reduced using the methods described by \cite{howell2011}. As shown in Extended Data Figure 3, the `Alopeke data rule out stellar companions within 1.2" within $\sim$5 mag at 562 nm and $\sim$5.5 mag at 832 nm at most angular separations. \\

\textbf{Stellar Atmospheric Parameter Determination}\\
The FIRE spectrum of TOI-6894 is shown in Extended Data Figure 5.
We used the SpeX Prism Library Analysis Toolkit \cite[SPLAT,][]{splat} to compare the spectrum to single-star spectral standards in the IRTF Spectral Library \cite{Cushing2005, Rayner2009}.
We find the best match to the M5 standard Wolf\,47, and thus we adopt a spectral type of M5.0 $\pm$ 0.5 for TOI-6894.
Following the approach of ref.\,\cite{Delrez2022}, we used the  relation between the equivalent widths of the $K$-band Na\,\textsc{i} and Ca\,\textsc{i} doublets and the H2O--K2 index \cite{Rojas-Ayala2012} to estimate the stellar metallicity \cite{Mann2014}.
This analysis yields a super-solar iron abundance estimate of $\mathrm{[Fe/H]} = +0.240 \pm 0.081$.

An independent spectral analysis was performed on the ESPRESSO spectra using ODUSSEAS \cite{ODUSSEAS2020}, a machine learning based code which has been specifically designed for performing spectral analysis of M-dwarf stars \cite[e.g.][]{lillobox2020lhs1140}. From this analysis we obtained values of \feh\,= \NfehODUSSEAS\ and \teff\,= \NteffODUSSEAS\,K. \\

\textbf{Global Analysis}\\
A joint analysis was performed to derive and constrain the stellar and planetary parameters of the \Nstar\ system. For this analysis, we used all available data: the TESS transit discovery photometry and all the follow-up photometry, the ESPRESSO and SPIRou radial velocity measurements, broad band photometry and astrometric data \cite[e.g. from][]{GAIADR32021}. The analysis followed the methods of \cite{hartman2019hats60_69}, \cite{bakos2020hats71} and \cite{hartman2023} and we direct the reader to those works for a more in depth discussion. We present the key details of the analysis here. Mandel and Agol transit models \cite{mandelagol2002transitmodel} were used to model the transit light curves. During the analysis the limb-darkening coefficients were fit as free parameters for each filter included, using Gaussian priors obtained from theoretical models \cite{claret2012LD, claret2013LD, claret2018LD}. A Keplerian orbit was assumed for modelling the RV measurements. 

Broadband photometry from \textit{Gaia}, 2MASS, and WISE was included in the analysis to constrain the stellar parameters along with the parallax measurement from \textit{Gaia} DR3 and stellar atmospheric parameters derived from the spectral analysis of the ESPRESSO and FIRE spectra. From these analyses we adopt Gaussian priors of $\feh= +0.240 \pm 0.081$ and \teff\,= \NteffODUSSEAS\,K for the joint analysis. We adopt the FIRE derived \feh\ value as the FIRE NIR spectrum provides a better S/N spectrum to determine the metallicity. However, we note that we ran an independent analysis taking the ESPRESSO derived metallicity as the prior range. The stellar and planetary parameters this independent analysis yielded, including the derived stellar metallicity, are fully consistent with those reported in this manuscript. At each step of the analysis the physical parameters of the host star were required to be consistent with the MIST stellar evolution models \cite[version 1.2; ][]{paxton2011mesa1,paxton2013mesa2,paxton2015mesa3,choi2016mist1,dotter2016mist2}, allowing for systematic errors in these models following the methods of \cite{hartman2023}. 

A differential evolution Markov Chain Monte Carlo procedure was used to fit the observations, using priors on the free parameters as listed in Supplementary Tables~1 and~2 (see also the discussion in \cite{hartman2019hats60_69}). After we performed an initial fit to the data, we applied a sigma clipping to the light curves to remove significant outliers, and we rescaled the uncertainties to give $\chi^2 / DoF = 1$ for each light curve. We then performed a second fit. For the majority of the light curves in this work this does not make a significant difference to the results of the fit. The exception is the MuSCAT2 light curves, which displayed some large outliers, likely due to clouds impacting the observations. These outliers were removed prior to the final analysis. The planetary and stellar parameters reported in this work represent the median and 1$\sigma$ uncertainty bounds calculated from the posterior distributions; these parameters are provided in Tables \ref{tab:stellar} and \ref{tab:planet}. 

From our analysis we found \Nplanet\ to be a transiting giant planet with a radius of $\rpl = \Nradius\,\rjup (\Nradiusearth\,\rearth)$ and a mass of $\mpl = \Nmass\,\mjup (\Nmassearth\,\mearth)$ which orbits its host star with an orbital period of $P = \Nperiod\,d$, semi-major axis of $a = \Nau$\,au, and an orbital eccentricity of \Necc. The 95\% upper limit placed on the orbital eccentricity is \NeccUpperLim. We note that the eccentricity value we measured is very close to zero, and consistent with zero within the errors. Therefore we are unable to constrain the argument of periastron of the orbit, $\omega$. Due to the low eccentricity value, we also repeated the analysis fixing the orbit to be circular. We compare the Bayesian Information Criterion (BIC) of the two models, computed using only the RV data points the light curve data does not contribute to constraining the eccentricity. We find a lower BIC for the eccentric model, but with a difference of just 0.6, suggesting that including the eccentricity as a free parameter is not strongly favoured by the data. However, we note that the measured planet and stellar parameters are fully consistent between the two models, with the free eccentricity model yielding slightly larger, more conservative uncertainties. As such we choose to report the parameters from the free eccentricity model in this paper.

We also found the host star \Nstar\ to be a very-low mass star with a mass and radius of $\mstar = \Nstarmass\,\msun$ and $\rstar = \Nstarradius\,\rsun$ and an effective temperature of $\teff = \Nteff\,K$. This makes \Nstar\ the lowest-mass star known to date to host a transiting giant planet, and just the fourth lowest-mass to host any transiting planet. We compare the host star to other low-mass stars that host transiting planets in Supplementary Figure 2. \\

\textbf{Blend Analysis}\\
In order to rule out the possibility that TOI~6894 is a blended stellar eclipsing binary system, we performed a blend analysis of the available observations following the method of \cite{hartman2019hats60_69}. Here we attempt to model the light curves, broad-band catalog photometry, spectroscopic atmospheric parameters, and astrometric parallax of the object as a blend between a bright M dwarf star and a fainter stellar eclipsing binary system. The parameters of the stars are constrained to follow the same MIST stellar evolution models used in the global joint analysis of the system.  We find that a blended stellar eclipsing binary scenario is easily ruled out in favor of a single star with a transiting planet, with $\Delta \chi^{2} = 1600$ between the best-fit blended eclipsing binary model and the best-fit transiting planet model.  We also rule out models consisting of an M dwarf star with a transiting giant planet and a non-transiting, fainter M dwarf companion with a mass down to the $0.1$\,\msun\ minimum stellar mass included in the MIST models.  In this case we find $\Delta \chi^2 = 140$ between the best-fit model with an unresolved stellar companion, and the best-fit model for a single star with a transiting planet. We note that all blend models considered have more free parameters than the single-star plus planet model, and thus are strongly disfavored by any model selection criteria. Therefore, from this analysis we can confidently rule out any blend scenarios and can be confident that the TOI-6894 system is a single star with a transiting planet companion. \\

\textbf{Chromaticity Analysis}\\
We perform a further analysis to investigate whether the transit depth of \Nplanet\ varies significantly with the wavelength of light in which the transit is observed. Such a chromatic variation would be evidence that the eclipse signals were due to a blended eclipsing binary, or could also point towards the presence of an unseen stellar companion. For this analysis, we perform a transit fit to each individual ground-based transit light curve obtained. We also perform a fit to each TESS sector individually. For these additional transit analyses, we only allow the planet-to-star radius ratio, \rprs, the quadratic limb-darkening coefficients, and the out-of-transit flux baseline to vary. The remaining transit parameters -- \tc, $P$, \ars, $i$ -- were fixed to the best fit results from the global analysis. For the radius ratio, \rprs, we use an uninformative uniform prior between 0 and 1, and for the limb darkening coefficients we use a wide Gaussian prior with the best fit result from the global analysis as the mean and five times the uncertainty as the standard deviation. We plot the results from this analysis in Extended Data Figure 2. From this analysis we find no evidence of a chromatic variation of the transit depth. \\

\textbf{Planet Composition Analysis}\\
We model the interior structure of \Nplanet\ within a retrieval framework for warm giant planets \cite{thorngren2019retrievals} which uses the forward models presented in \cite{thorngren2016forwardmodel}. From this analysis we find a metal mass fraction -- i.e. the fraction of the planet mass which is not hydrogen or helium -- of $Z_{\rm P} = \NZp$. We note that the uncertainty quoted here is a statistical error based on the uncertainties on the stellar mass, radius, and age. Combining with the overall mass of the planet we find a metal mass content of \Nplanet\ of $M_{\rm metal} = \Nsolidmass\,\mearth$. 

An empirical mass-radius relation for cool giant planets was derived using the known population in \cite{thorngren2019massradrelation}. From this known population, its mass-radius relation, and the resulting dispersion in the population, the median planet radius for a $0.164\,\mjup$ planet is $0.67\,\rjup$ with a $1\,\sigma$ dispersion of $\pm 0.18\,\rjup$. Therefore, we can see while \Nplanet\ is lower density than the median planet expected from the bulk population, its radius of \rpl = \Nradius\,\rjup\ is consistent with the dispersion seen in the overall population to within a tolerance of $1\,\sigma$. \\

\textbf{Search for Additional Planets and Detection limits}\\
We analyzed the TESS 120s data with the \texttt{SHERLOCK} package \cite{pozuelos2020,demory2020}. We refer the readers to \cite{pozuelos2023} and \cite{devora2024} to learn about recent usages and searching strategies. We first found a strong signal corresponding to the known 3.37\,d planet, which enabled us to confirm independently to the SPOC pipeline the detectability of this planet in TESS data. We found no other signal hinting at extra transiting planets in orbital periods ranging from 0.5 to 15\,d.  

We then performed injection and retrieval experiments on this dataset to establish detection limits. We employed the \texttt{MATRIX} package \cite{devora2023}, which generates a sample of synthetic planets by combining a range of orbital periods, planetary radii, and orbital phases injected in the data. 

In particular, we generated 36,000 scenarios and searched them for transit-like features mimicking the procedure conducted by \texttt{SHERLOCK}. From the results displayed in Supplementary Figure 3, we conclude that TESS data allows us to detect large transiting planets (R$>$5~R$_\oplus$) in short orbital periods (P$<$6\,d) easily, with recovery rates of $\sim$100\%. Indeed, TOI-6894\,b falls in this region. These planets become more challenging to detect for longer orbital periods but still possible, with recovery rates between 40\% and 80\%.
These results allowed us to conclude that the existence of these planets in the system is very unlikely. On the other hand, small transiting planets with sizes smaller than 4 R$_{\oplus}$ would be undetectable in the complete set of periods explored. Hence, we can not offer any constraint on the existence of these planets in the system. \\

\textbf{Atmospheric Characterisation Prospects}\\
We expect TOI-6894\,b to become a benchmark planet for the study of temperate H/He atmospheres. TOI-6894\,b receives a stellar irradiation $S = \NinsolEarth \rm \, S_\oplus$, that translates into an equilibrium temperature $T_{\rm eq} = \NTeqA \,\rm K$. This value assumes an albedo $A = 0.1$, similar to many hot and warm Jupiters \cite{wong2020}. At this temperature, it is widely expected the planet will be dominated by methane chemistry \cite[similarly to WASP-80\,b][]{triaud2015,bell2023}. Using these properties, we model likely atmospheres with and without clouds, and high and low C/O ratios and find that methane absorption features in the planet's transmission spectrum are expected to have amplitudes of $6000$, $9000$, and $11000~\rm ppm$ in the optical, near infrared and mid-infrared, well in excess of any other giant to date, particularly for planets with a similarly low equilibrium temperature. This is mainly caused by two effects. The host star, \Nstar, is small and transmission features are amplified by $\rstar^{-2}$, and by the surprising low surface gravity of \Nplanet. To address the detectability of individual molecules within the TOI-6894b planet specra, we used the methodology applied in \cite{chouqar2020properties}, \cite{lustig2019detectability} and \cite{morley2017observing} for transit geometry. We run the PANDEXO JWST noise model across a grid in number of transits from 1 to 100, which is sufficient to establish a simple S/N scaling relationship and we determine the S/N on the difference between the model spectrum and the fiducial spectrum. Our Pandexo  simulations of observations using the NIRISS/SOSS, NIRSpec/G395M, and MIRI/LRS modes on TOI-6894b showed that a single transit could suffice to retrieve abundances of key atmospheric species like methane, water, and carbon dioxide with a total expected S/N $\geq $100. We plot example transmission spectra obtained from PANDEXO in Extended Data Figure 8. Furthermore, as illustrated in Extended Data Figure 8, molecular absorption features should be detectable at wavelengths beyond 2 microns,  even with a cloud deck at 1 mbar. To place this planet in context we calculated the transmission spectroscopy metric \cite[TSM; see ][ for more details]{kempton2018tsm} which can be used as a measure of how amenable a planet is to atmospheric characterisation through transmission spectroscopy. We find \Nplanet\ to have a TSM value of $356 \pm 58$. Comparing the TSM of \Nplanet\ to the TSM values for other known planets (Figure 3 and Extended Data Figure 7), we find \Nplanet\ to have the highest TSM value of any giant planet with a host star less massive than 0.7\,\msun\ and the second highest for any planet with a low-mass host star ($\mstar \leq 0.4 \msun$), second only to GJ~1214\,b. \Nplanet\ particularly stands out when considering other planets with a low equilibrium temperature. Assuming the planet has an albedo, $A = 0.1$, we also expect its emission spectrum to be highly amenable for detection of atmospheric features. Like for transmission, we model possible emission spectra and find typical eclipse depths of $1000 - 6000~\rm ppm$ in the mid-infrared. Studying the atmosphere of \Nplanet\ can provide an easy access to an H/He atmosphere intermediate between those of hot Jupiters and our Solar system's Jupiter and studying its chemistry will contribute to refining atmospheric models. In addition, studying a planet's atmosphere might bring further clues related to its formation history. The star is metal-rich ($\feh = \Nmetal$), and it will be of interest to measure whether the planet's atmosphere is as well. Such measurements would reveal the true metal content of \Nplanet, thereby also revealing the composition of \Nplanet\ and clues into its formation history \cite{blootmiguel2023}.

\section{Extended Data Tables and Figures}
\begin{figure}
    \centering
    \includegraphics[width=0.75\textwidth]{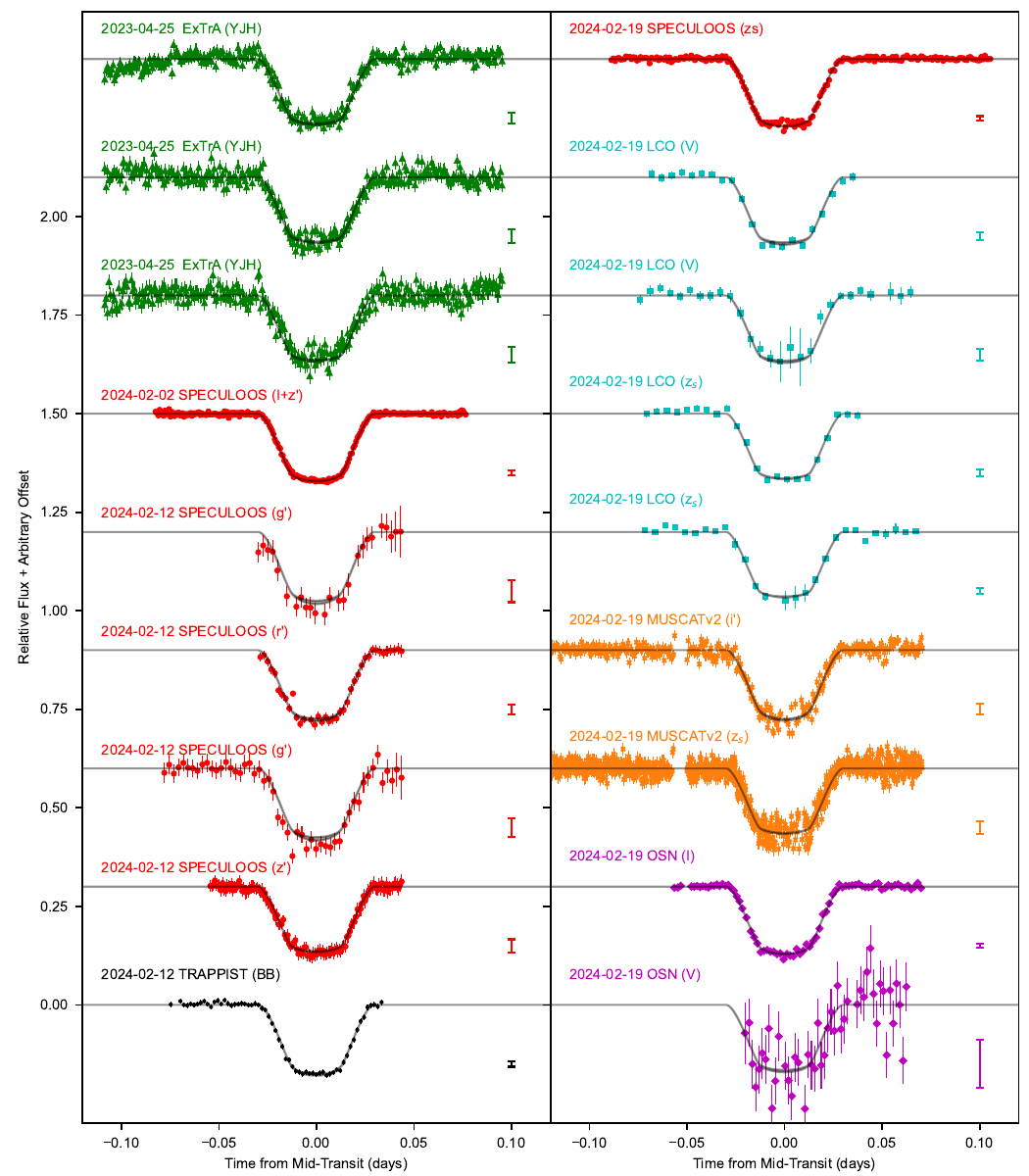}
    \caption{\textbf{Extended Data Figure 1: }Ground-based light curve transit observations for \Nplanet. Each light curve is plotted individually and all light curves are offset from one another for clarity. The labels of each light curve give the date on which the observations were performed and the filter used for the observations. The different colours and markers denote the telescope used to obtain the observations: ExTrA (green triangles); SPECULOOS (red circles); TRAPPIST (black points); LCO (cyan squares); MUSCAT2 (orange crosses); OSN (purple diamonds). The the gray shaded regions provide the $1\sigma$ confidence region for the transit models. The errorbars provided are the reported uncertainties for all light curves except the MuSCAT2 data, for which the reported uncertainties were significantly over-estimated and so we plot the rescaled uncertainties (see Methods) for clarity. For some observations the uncertainties are too small to be seen, and so we provide the median uncertainty for all observations as the errorbars plotted to the right of the corresponding observations.}
    \label{fig:tfop_lcs}
\end{figure}

\begin{figure}
    \centering
    \includegraphics[width=0.9\textwidth]{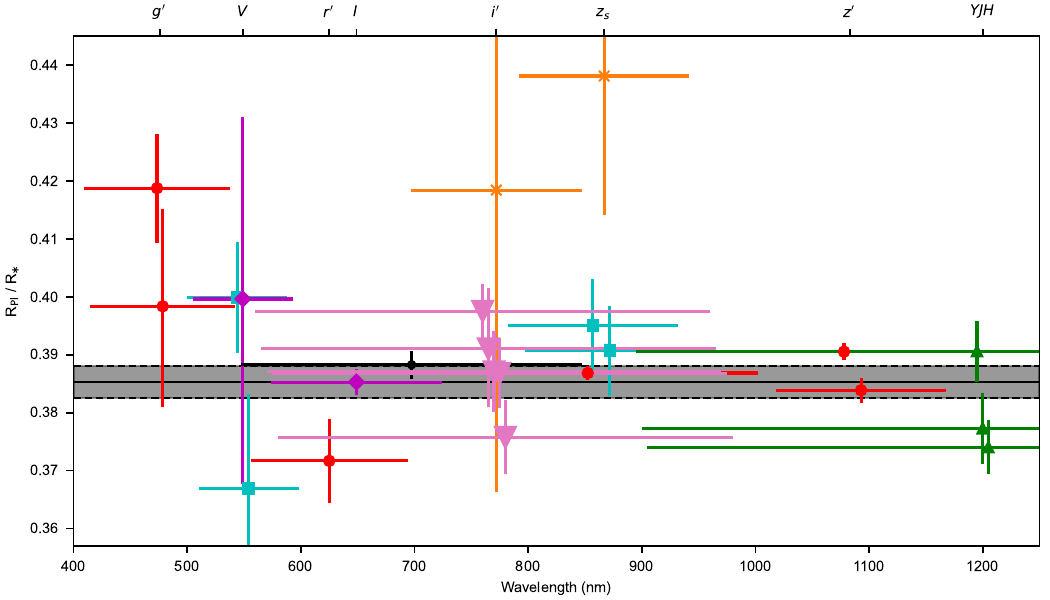}
    \caption{\textbf{Extended Data Figure 2: }Individual \rprs\ values obtained from transit fit analyses performed for each TESS sector and each individual ground-based transit light curve obtained. The x-axis plots the reference wavelength for each filter used. Where more than one result uses the same filter the points are offset slightly in the x-direction for clarity. The y-axis errorbars give the 1$\sigma$ uncertainty from the transit analysis and the x-axis errorbars show the FWHM of the observing filter used. The upper axis shows the reference wavelengths for some of the filters used. The data point markers and colours are the same as for Extended Data Figure 1, with the addition of the pink downward triangles for the TESS results. The solid black line and shaded grey region give the best fit \rprs\ value and 1$\sigma$ uncertainty from the global analysis.}
    \label{fig:chrom}
\end{figure}

\begin{figure*}
    \centering
    \includegraphics[width=0.45\textwidth]{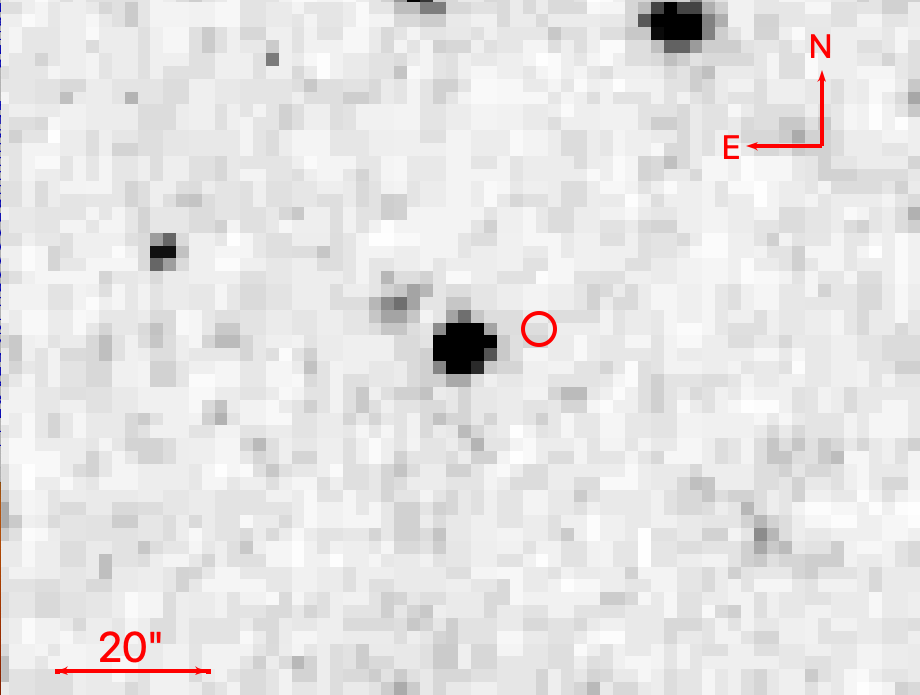}
    \includegraphics[width=0.45\textwidth]{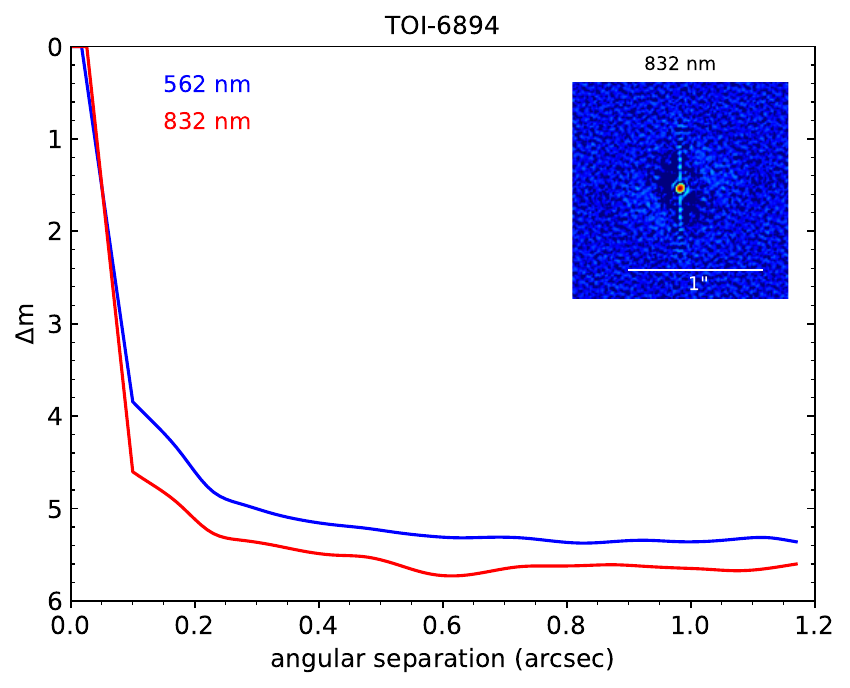}
    \caption{\textbf{Extended Data Figure 3: }Observations to check for signs of blended companions. \textbf{Left: }The R-band 1-hour exposure time DSS image from the 48-Inch Palomar telescope observed in 1952 January 31.  \Nstar\ is the star at the center of the image.  The current sky location from Gaia DR3 is shown with a red circle. 
 No background source is detected to the limit of the DSS plate ($G=19.5$). \textbf{Right: }Contrast curves obtained using the `Alopeke speckle imager at Gemini North. The inset shows the obtained speckle image.} 
    \label{fig:dss}
\end{figure*}

\begin{figure}
    \centering
    \includegraphics[width=0.8\linewidth]{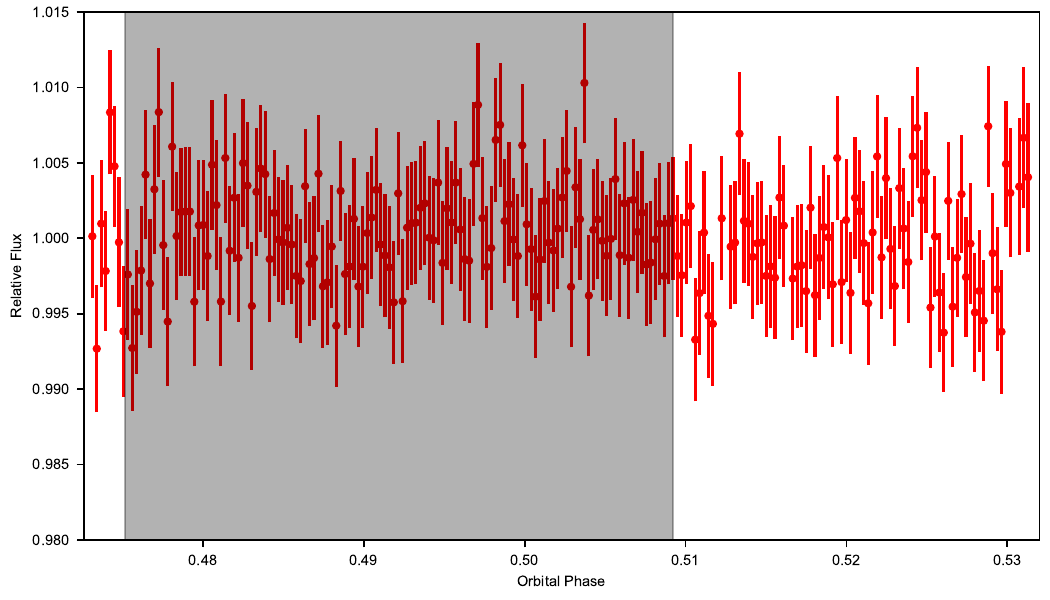}
    \caption{\textbf{Extended Data Figure 4: }SPECULOOS observations taken during the prediction time of a secondary eclipse, with the error bars showing the reported photometric uncertainties. The gray shaded region gives the 1$\sigma$ window for the estimated time of the occultation, accounting for the eccentricity posterior distribution. No significant secondary eclipse is observed.}
    \label{fig:spec_occult}
\end{figure}

\begin{figure}
    \centering
    \includegraphics[width=0.5\linewidth]{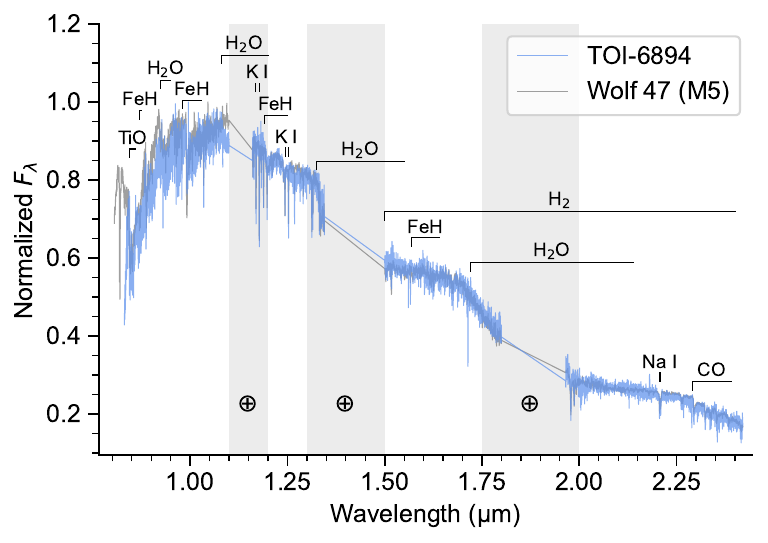}
    \caption{\textbf{Extended Data Figure 5: }
    Magellan/FIRE spectrum of TOI-6894.
    The target spectrum (blue) is shown along with the spectrum of the best-fit M5 standard (grey).
    Regions of strong telluric absorption are shaded in grey, and prominent atomic and molecular features of M dwarfs are highlighted.
    \label{fig:fire}
    }
\end{figure}

\begin{figure}
    \centering
    \includegraphics[width=0.9\linewidth]{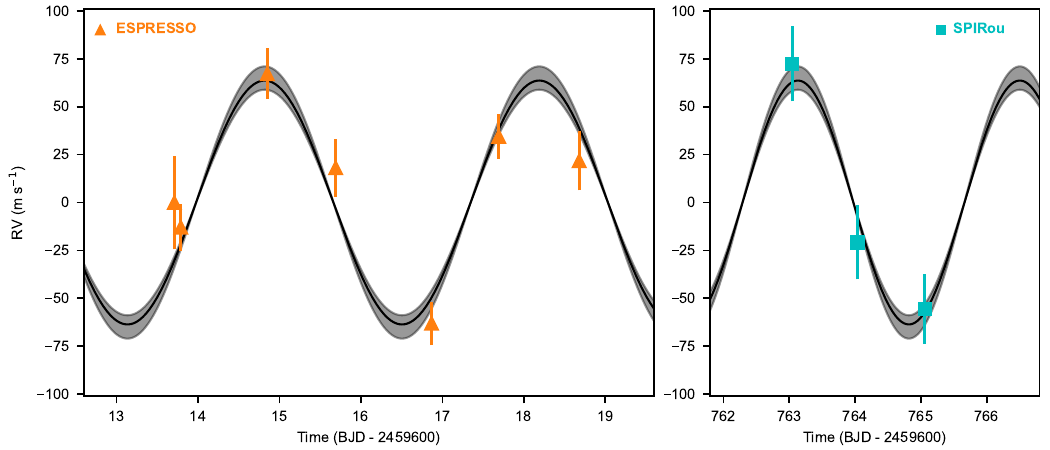}
    \caption{\textbf{Extended data Figure 6: }Radial velocity time series for \Nstar. The ESPRESSO data is plotted in the left panel and the SPIRou data in the right, and the systemic radial velocity values (see Extended Data Table 2) have been subtracted from the respective radial velocity time series. The errorbars provided the 1$\sigma$ radial velocity uncertainties yielded by the reduction pipelines. The symbols, colours, and model lines are the same as presented in Figure~1b.}
    \label{fig:rv_timeseries}
\end{figure}

\begin{figure}
    \centering
        \includegraphics[width=0.9\linewidth]{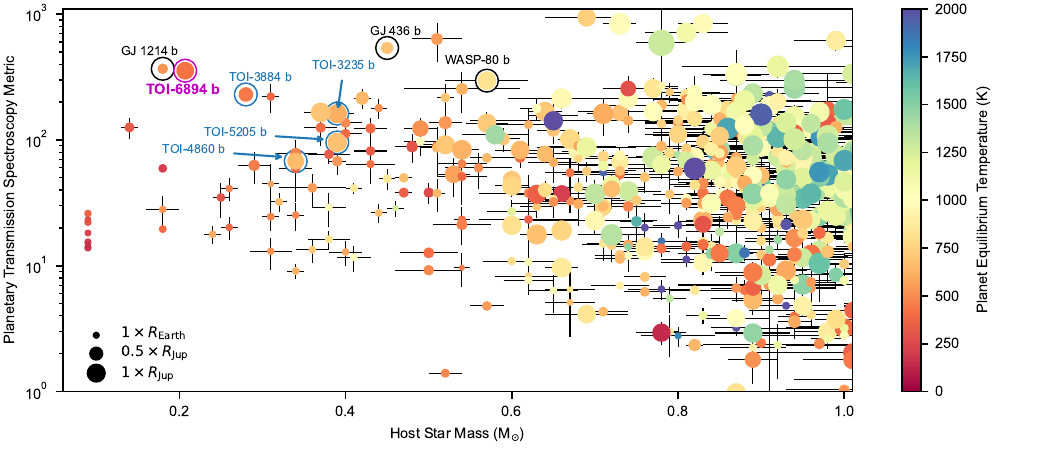}
    \caption{\textbf{Extended Data Figure 7: }Transmission spectroscopy metric (TSM) as a function of host star mass for transiting exoplanets with mass measurements. Data and uncertainties are extracted from the NASA Exoplanet Archive, with the errorbars showing the reported 1$\sigma$ uncertainties on each parameter. The points are coloured according to their planetary equilibrium temperature. The size of the points scale with the planetary radius. TOI-6894\,b is highlighted by the magenta circle. We also highlight transiting giant and sub-giant planets with mid-M-dwarf host stars with the blue circles and arrows and other planets with high TSM already observed and/or scheduled on JWST with the black circles.}
    \label{fig:TSM}
\end{figure}

\begin{figure*}
	\centering
	\includegraphics[width=0.45\textwidth]{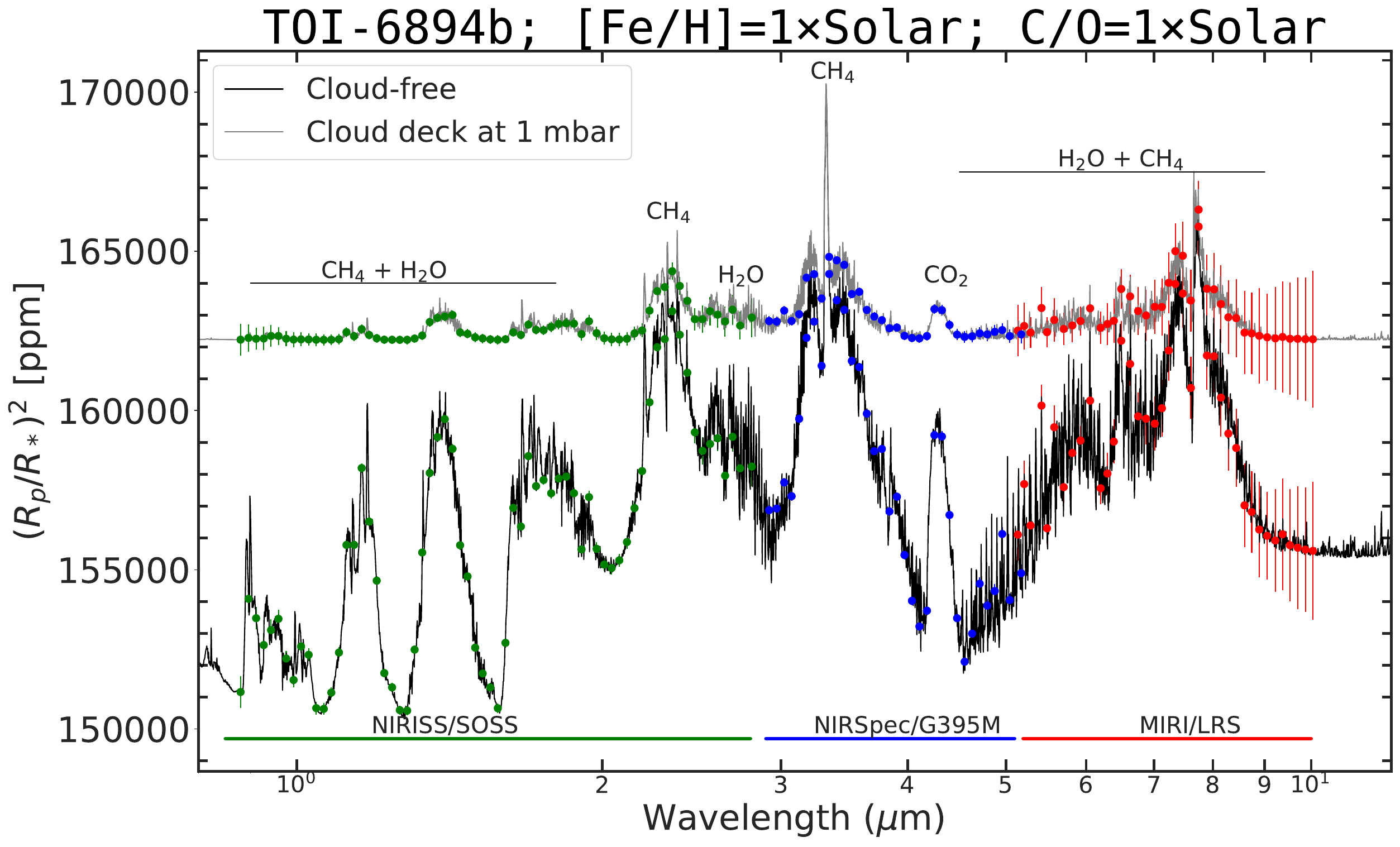}
        \includegraphics[width=0.45\textwidth]{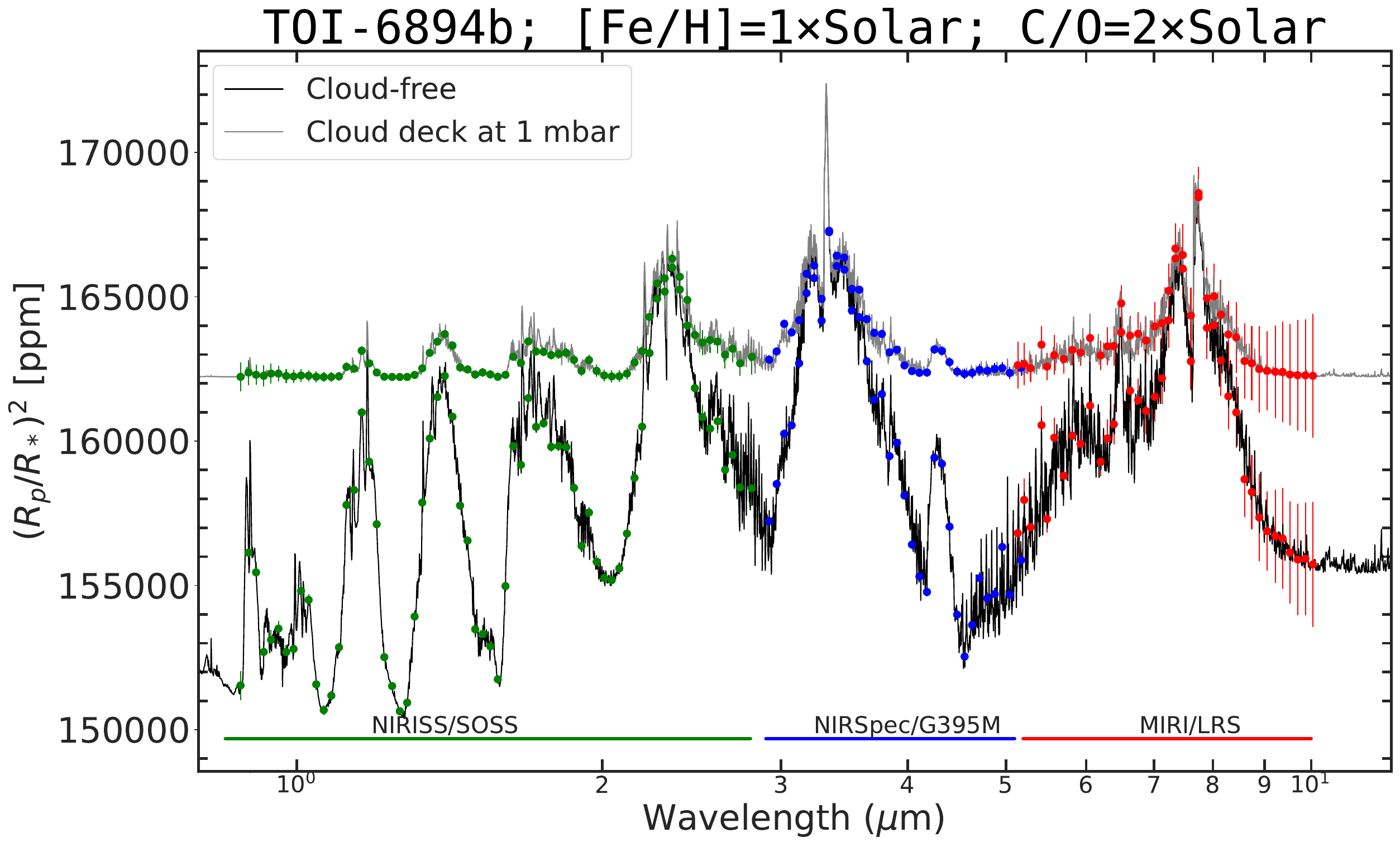} 
\caption{\textbf{Extended Data Figure 8: }PandExo simulated transmission spectra of \Nplanet. Left panel: Clear and cloudy transmission spectra models assuming solar abundance are shown as solid lines. Right panel: Clear and cloudy models with the carbon-to-oxygen ratio enhanced by a factor of two. PandExo simulated observations with 1 transit for JWST NIRISS-SOSS, NIRSpec-G395M, and MIRI-LRS modes are also depicted, with their wavelength coverage indicated by coloured solid lines. The errorbars for both panels provide the estimated 1$\sigma$ measurement uncertainties provided by PandExo.}
	\label{spec-tr}		
\end{figure*}

{\centering
\textbf{Extended Data Table 1: }Ground-based follow-up observations log for TOI-6894.01.
{\renewcommand{\arraystretch}{1.2}
\begin{tabular}{lccccccc}
\toprule
\vspace{0.15cm}
Telescope  & Filter & Date &  Exptime (s) & FWHM (\arcsec) & Aperture (\arcsec) & Coverage \\
\toprule
 \multicolumn{7}{c}{\bf Transit} \\
 \hline
ExTrA & $YJH$ & 2023 Apr 25 & 60 & 1.4 & 4 & Full \\
SPECULOOS-S/Europa & $I+z'$ & 2024 Feb 02  & 49  & 1.2 & 1.7 & Full  \\
SPECULOOS-S/Io & Sloan-$g'$ & 2024 Feb 12  &  200 & 2.0 & 1.9 & Full  \\
SPECULOOS-S/Europa & Sloan-$r'$ & 2024 Feb 12  &  150 & 1.4 & 1.6 & Full  \\
SPECULOOS-S/Ganymede & Sloan-$g'$ & 2024 Feb 12  & 200  & 1.6 & 1.5 & Full  \\
SPECULOOS-S/Callisto & Sloan-$z'$ & 2024 Feb 12  & 60  & 2.5 & 1.7 & Full  \\
TRAPPIST-S & $BB$ & 2024 Feb 12  & 140  & 2.1 & 3.7 & Full  \\
SPECULOOS-N/Artemis & Sloan-$z'$ & 2024 Feb 19  &  70 & 1.2 & 1.4 & Full  \\
LCOGT/SAAO & $V$ & 2024 Feb 19 & 300 & 2.4 & 3.1 & Full \\
LCOGT/Teide & $V$ & 2024 Feb 19 & 300 & 1.7 & 1.9 & Full \\
LCOGT/SAAO & $z_s$ & 2024 Feb 19 & 70 & 2.1 & 2.7 & Full \\
LCOGT/Teide & $z_s$ & 2024 Feb 19 & 70 & 1.5 & 2.3 & Full \\
TCS/MuSCAT2 & $i'$,$z_s$ & 2024 Feb 19  &  45, 15 & 3.0, 2.9 & 10.9, 10.9 & Full  \\
OSN/T150 & $I$ & 2024 Feb 19  &  120 & 2.3 & 4.6 & Full  \\
OSN/T150 & $V$ & 2024 Feb 19  &  90 & 2.5 & 3.7 & Egress  \\
 \hline
 \multicolumn{7}{c}{\bf Occultation} \\
 \hline
 SPECULOOS-S/Europa & Sloan-$z'$ & 2024 Feb 07  & 70  & 1.1 & 1.9 & Full  \\
\hline
\end{tabular}}

\textbf{Extended Data Table 2: }Radial Velocity information for \Nstar. We provide both the measured radial velocities and their uncertainties, as well as the systemic RV and jitter values determined for each instrument.

    \begin{tabular}{c|c|c|c}
    \toprule
    \vspace{0.15cm}
    \textbf{Time} & \textbf{Radial Velocity} & \textbf{Error} & \textbf{Instrument} \\
    BJD TDB & \ms & \ms & \\
    \toprule
    \hline
    2459613.709421 & 15826.34 & 24.38 & ESPRESSO \\ 
    2459613.789054 & 15813.43 & 12.21 & ESPRESSO \\ 
    2459614.851345 & 15893.57 & 13.24 & ESPRESSO \\ 
    2459615.690120 & 15844.47 & 15.19 & ESPRESSO \\ 
    2459616.867351 & 15763.30 & 11.30 & ESPRESSO \\ 
    2459617.692834 & 15860.84 & 11.56 & ESPRESSO \\ 
    2459618.680050 & 15848.15 & 15.44 & ESPRESSO \\
    2460363.044 &16042.33 &19.60 & SPIRou \\
    2460364.036 &15949.21 &19.33 & SPIRou \\
    2460365.059 &15914.44 &18.15 & SPIRou \\
    \toprule
    \textbf{Parameter} & \textbf{Symbol} & \textbf{unit} & \textbf{Value} \\
    \toprule
    ESPRESSO Systemic RV & $\gamma_{\rm RV; ESPRESSO}$ & \ms & \NgammaESP \\
    ESPRESSO RV Jitter & $\sigma_{\rm RV; ESPRESSO}$ & \ms & \NjitterESP \\
    SPIRou Systemic RV & $\gamma_{\rm RV; SPIRou}$ & \ms & \NgammaSpirou \\
    SPIRou RV Jitter & $\sigma_{\rm RV; SPIRou}$ & \ms & \NjitterSpirou \\
    \toprule   
    \end{tabular}
}

\section{Supplementary Information}
\begin{figure}
    \centering
    \includegraphics[width=0.7\textwidth]{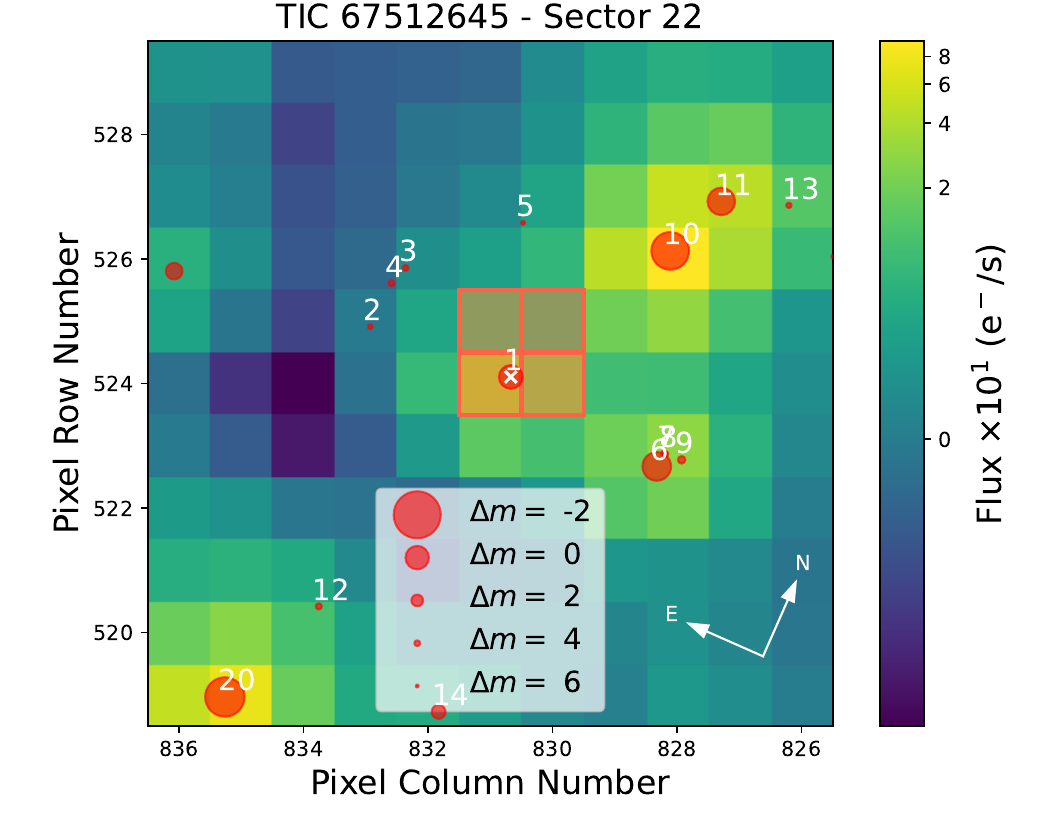}
    \caption{\textbf{Supplementary Figure 1: }A $11\times 11$ pixel cutout of the TESS image aronud the location of \Nstar\ from Sector~22, plotted using \textsc{tpfplotter} \cite{aller2020tpfplotter}. The red shaded boxes highlight the aperture used by the SPOC pipeline to extract the photometric light curve, and the nearby sources from \textit{Gaia} are labeled with the red circles. The size of each marker corresponds to the magnitude of the star in the \textit{Gaia G} band relative to \Nstar.}
    \label{fig:piximage}
\end{figure}

\begin{figure}
    \centering
    \includegraphics[width=0.8\textwidth]{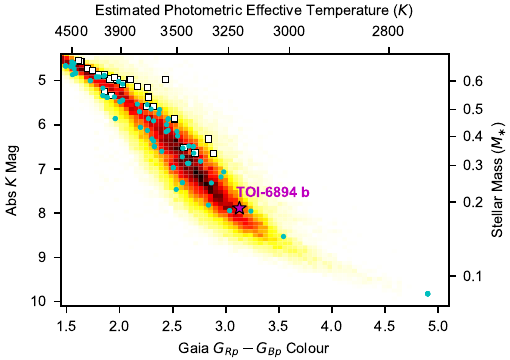}
    \caption{\textbf{Supplementary Figure 2: }Colour-magnitude diagram displaying the \textit{Gaia} $G_{\rm Bp} - G_{\rm Rp}$ colour and absolute \textit{2MASS} $K$ magnitude. The 2D histogram heat map shows the distribution of the population of low-mass stars studied by \cite{bryant2023lmstargiantplanetoccrates}. The individual markers show the host stars of \Nplanet\ (purple star); known transiting giant planets (black open squares\textbf{; \mpl$ \geq 0.1$\,\mjup}); and other known transiting planets (cyan circles). The parameters of known planet host stars are taken from the NASA Exoplanet Archive (accessed 16 May 2024). The upper axis provides an approximate representation of the stellar effective temperature, using the scaling provided in \cite{mann2015constrainMDwarf1}, and the right-hand axis provides a representation of the stellar mass, computed using the scaling from \cite{mann2019constrainMDwarf2}.}
    \label{fig:hrd}
\end{figure}

\begin{figure}
    \centering
    \includegraphics[width=0.8\textwidth]{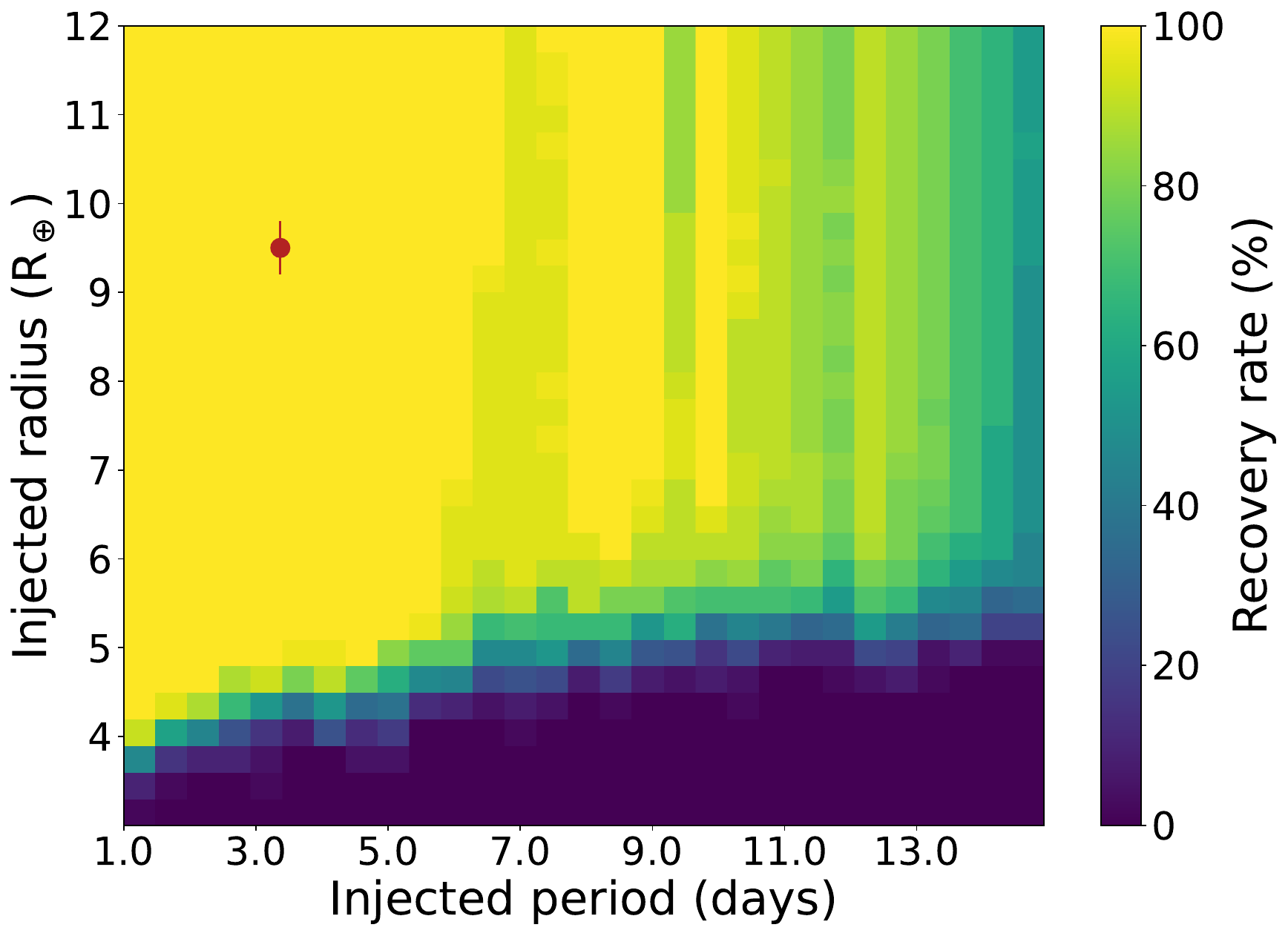}
    \caption{\textbf{Supplementary Figure 3: }Injection-and-retrieval experiment conducted to test the detectability of extra planets in the system TOI-6894 using the TESS 120\,s data, corresponding to Sector 72. We explored a total of 36000 different scenarios. Each pixel shows the evaluation of about 40 scenarios, that is, 40 light curves with injected planets having different $P_{\mathrm{planet}}$, $R_{\mathrm{planet}}$, and T$_{0}$. Larger recovery rates are presented in yellow and green colors, while lower recovery rates are shown in blue and darker hues. Planets smaller than 4.0~R$_{\oplus}$ would be undetectable for the explored periods. The red dot refers to the planet TOI-6894\,b.}
    \label{fig:inject}
\end{figure}

\newcommand{\specialcell}[2][c]{%
  \begin{tabular}[#1]{@{}c@{}}#2\end{tabular}}

\begin{table}
    \centering
    \caption{\textbf{Supplementary Table 1: }Physical Parameters varied in joint analysis}
{\renewcommand{\arraystretch}{0.8}
\begin{tabular}{lcc} 
    \hline
    \textbf{Parameter}	&	\textbf{Prior}		& \textbf{Notes}\\
    \hline
    $T_{A}$ & uniform & mid transit time of first observed transit \\
$T_{B}$ & uniform & mid transit time of last observed transit \\
$K$ & uniform, $K > 0$& RV semi-amplitude \\
$\sqrt{e}\cos\omega$ & uniform, $0 \leq e < 1$ & eccentricity parameter, either fixed to zero or varied \\
$\sqrt{e}\sin\omega$ & uniform, $0 \leq e < 1$ & eccentricity parameter, either fixed to zero or varied \\
$\rpl/\rstar$ & uniform & ratio of planetary to stellar radius \\
$b^{2}$ & uniform, $b^{2} \geq 0$ & impact parameter squared \\
$\zeta/\rstar$ & uniform & reciprocal of the half duration of the transit \\
$\gamma_{i}$ & uniform & systemic velocity for RV instrument $i$ \\
$LD_{b,j}$ & \specialcell{Gaussian with $\sigma = 0.2$\\ mean based on \cite{claret2012LD,claret2013LD,claret2018LD}} & Linear limb darkening coefficient for filter $j$ \\
$LD_{b,j}$ & \specialcell{Gaussian with $\sigma = 0.2$\\ mean based on \cite{claret2012LD,claret2013LD,claret2018LD}} & Quadratic limb darkening coefficient for filter $j$ \\
$d_{\rm mod}$ & $2\ln(\frac{d_{\rm mod}+5}{5})-\frac{d_{\rm mod}+5}{7650}$ & \specialcell{distance modulus, note the {\em Gaia} DR3 parallax \\ is treated as an observable to be fit} \\
$A_{V}$ & \specialcell{Gaussian with $\sigma=0.25$\,mag\\mean based on MWDUST model\\ $A_{V} \geq 0$} & extinction \\
$T_{\rm eff}$ & Gaussian, \NteffODUSSEAS, $T_{\rm eff} > 0$ & host star effective temperature \\
\feh & Gaussian, $+0.240 \pm 0.081$ & host star metallicity \\
\hline
 \end{tabular}}
    \label{tab:parametersvaried}
\end{table}

\newpage

\begin{table}
    \centering
    \caption{\textbf{Supplementary Table 2: }Auxiliary Parameters varied in joint analysis}
{\renewcommand{\arraystretch}{0.8}
\begin{tabular}{lcc} 
    \hline
    \textbf{Parameter}	&	\textbf{Prior}		& \textbf{Notes}\\
    \hline
$\sigma_{\rm jit,i}$ & $-\log(\sigma_{\rm jit,i})$, $\sigma_{\rm jit,i} > 0$ & jitter for RV instrument $i$ \\
$m_{0,TESS,i}$ & uniform & out-of-transit magnitude for {\em TESS} light curve $i$ \\
$d_{TESS,i}$ & uniform, $0 < d_{HS,i} \leq 1$ & transit dilution factor for {\em TESS} light curve $i$ \\
$m_{0,LC,i}$ & uniform & out-of-transit magnitude for follow-up light curve $i$ \\
$m_{1,LC,i}$ & uniform & \specialcell{linear trend to out-of-transit magnitude for\\ follow-up light curve $i$} \\
$m_{2,LC,i}$ & uniform & \specialcell{quadratic trend to out-of-transit magnitude for\\ follow-up light curve $i$} \\
$\delta x_{0,LC,i}$ & uniform & \specialcell{linear detrending coefficient for CCD $\Delta x$\\ position of star for follow-up light curve $i$.\\Used for SPECULOOS, TRAPPIST, and OSN.} \\
$\delta y_{0,LC,i}$ & uniform & \specialcell{linear detrending coefficient for CCD $\Delta y$\\ position of star for follow-up light curve $i$.\\Used for SPECULOOS, TRAPPIST, and OSN.} \\
$fwhm_{0,LC,i}$ & uniform & \specialcell{linear detrending coefficient for FWHM of star\\ for follow-up light curve $i$.\\Used for SPECULOOS, TRAPPIST, and OSN.} \\
$sky_{0,LC,i}$ & uniform & \specialcell{linear detrending coefficient for sky background\\ for follow-up light curve $i$.\\Used for SPECULOOS, TRAPPIST, and OSN} \\
$\sigma_{\mstar,sys}$ & Gaussian with $\sigma = 5$\% & Fractional systematic uncertainty on $\mstar$\\
$\sigma_{\feh,sys}$ & Gaussian with $\sigma = 0.08$\,dex & Systematic uncertainty on $\feh$ \\
$\sigma_{\teff,sys}$ & Gaussian with $\sigma = 4$\% & Fractional systematic uncertainty on $\teff$\\
$\sigma_{M_{bol},sys}$ & Gaussian with $\sigma = 0.021$\,mag & Systematic uncertainty on bolometric magnitude \\
\hline
 \end{tabular}}
    \label{tab:parametersvaried2}
\end{table}

\section{Data Availability}
The TESS-SPOC FFI photometry we used is publicly available as a High-Level Science Product from the Mikulski Archive for Space Telescopes (MAST; \url{https://archive.stsci.edu/hlsp/tess-spoc}; \cite{TSFDOI}).
The ESPRESSO and SPIRou RV data is provided in Extended Data Table 2 within this paper. The ESPRESSO observations were obtained under ESO programme ID 108.22B4.001 and the raw spectra can be obtained from the ESO Science Portal under target name TIC67512645 (\url{https://archive.eso.org/scienceportal/home}). The Magellan/FIRE spectrum (Data Tag 441942) is available via the ExoFoP-TESS archive (\url{https://exofop.ipac.caltech.edu/tess/target.php?id=67512645}).
The ExTrA data (Data Tag 441923), SPECULOOS data (Data Tags 438216, 438351, and 438530), TRAPPIST data (Data Tag 438352), LCOGT data (Data Tag 438460), MuSCAT2 data (Data Tag 441940), and OSN data (Data Tag 441978) are available via the ExoFoP-TESS archive (\url{https://exofop.ipac.caltech.edu/tess/target.php?id=67512645}). 
The Gemini North speckle imaging data (Data Tag 441696) is available via the ExoFoP-TESS archive (\url{https://exofop.ipac.caltech.edu/tess/target.php?id=67512645}).

\section{Code Availability}
The code used to run the main MCMC analysis has been previously described in \cite{hartman2019hats60_69,bakos2020hats71,hartman2023}.
The SPLAT code is available from \url{https://github.com/aburgasser/splat} and the ODUSSEAS code is available from \url{https://github.com/AlexandrosAntoniadis/ODUSSEAS}.
The PROSE code is available \url{https://github.com/lgrcia/prose}; AstroImageJ is described in \cite{Collins2017} and available from \url{https://www.astro.louisville.edu/software/astroimagej/}; the BANZAI code is described in \cite{McCully2018} and available from \url{https://github.com/LCOGT/banzai}; the MuSCAT2 data reduction pipeline is described in \cite{Parviainen2020}; the FIRE bright source data reduction pipeline is described in \cite{2020ASPC..522..623C}; the ESPRESSO DRS pipeline is available from \url{https://www.eso.org/sci/software/pipelines/espresso/espresso-pipe-recipes.html}; the APERO pipeline is described in \cite{Cook2022} and is available from \url{https://github.com/njcuk9999/apero-drs}.

\section{Acknowledgements}
The contributions at the Mullard Space Science Laboratory by E.M.B. and V.V.E have been supported by UK's Science \& Technology Facilities Council through the STFC grants ST/W001136/1 and ST/S000216/1.
A.J.\ acknowledges support from ANID -- Millennium  Science  Initiative -- ICN12\_009 and from FONDECYT project 1251439.
J.D.H and G.A.B. acknowledge funding from NASA XRP grant 80NSSC22K0315.
This work is partly supported by the National Science Foundation of China (Grant No. 12133005).
We acknowledge financial support from the Agencia Estatal de Investigaci\'on of the Ministerio de Ciencia e Innovaci\'on MCIN/AEI/10.13039/501100011033 and the ERDF “A way of making Europe” through project PID2021-125627OB-C32, and from the Centre of Excellence “Severo Ochoa” award to the Instituto de Astrofisica de Canarias.
This work is partly supported by JSPS KAKENHI Grant Number JP24H00017 and JSPS Bilateral Program Number JPJSBP120249910.
F.J.P, P.J.A and V.C acknowledges financial support from the Agencia Estatal de Investigación (AEI/10.13039/501100011033) of the Ministerio de Ciencia e Innovación and the ERDF “A way of making Europe” through projects PID2022-137241NB-C43 and the Centre of Excellence “Severo Ochoa” award to the Instituto de Astrofísica de Andalucía (CEX2021-001131-S). 
This publication benefits from the support of the French Community of Belgium in the context of the FRIA Doctoral Grant awarded to M.T.
T.D. acknowledges support from the McDonnell Center for the Space Sciences at Washington University in St. Louis.
EJ is FNRS Senior Research Associate
H.P. acknowledges support by the Spanish Ministry of Science and Innovation with the Ramon y Cajal fellowship number RYC2021-031798-I and
funding from the University of La Laguna and the Spanish Ministry of Universities.
B.V.R. thanks the Heising-Simons Foundation for Support.
This material is based upon work supported by the National Aeronautics and Space Administration under Agreement No.\ 80NSSC21K0593 for the program ``Alien Earths''.
The results reported herein benefitted from collaborations and/or information exchange within NASA’s Nexus for Exoplanet System Science (NExSS) research coordination network sponsored by NASA’s Science Mission Directorate.
IC's research is supported by the Telescope Data Center, Smithsonian Astrophysical Observatory.
R.B. acknowledges support from FONDECYT Project 1241963 and from ANID -- Millennium  Science  Initiative -- ICN12\_009.
Participation of K.N. was made possible by the SETI Institute REU internship program (NSF award 2051007).
This research was carried out at the Jet Propulsion Laboratory, California Institute of Technology, under a contract with the National Aeronautics and Space Administration (80NM0018D0004).
This research was supported by Wallonia-Brussels International (WBI).
This research has made use of the NASA Exoplanet Archive, which is operated by the California Institute of Technology, under contract with the National Aeronautics and Space Administration under the Exoplanet Exploration Program.
We acknowledge funding from the European Research Council under the ERC Grant Agreement n. 337591-ExTrA. ExTrA has been supported by Labex OSUG@2020 (Investissements d'avenir -- ANR10 LABX56), the "Programme National de Physique Stellaire" (PNPS) and the “Programme National de Palnétologie of CNRS/INSU, co-funded by CEA and CNES.
This paper includes data gathered with the 6.5 meter Magellan Telescopes located at Las Campanas Observatory, Chile.
Based on observations made at the Observatorio de Sierra Nevada (OSN), operated by the Instituto de Astrofísica de Andalucía (IAA-CSIC).
Based on observations obtained at the Canada-France-Hawaii Telescope (CFHT) which is operated from the summit of Maunakea by the National Research Council of Canada, the Institut National des Sciences de l'Univers of the Centre National de la Recherche Scientifique of France, and the University of Hawaii. The observations at the Canada-France-Hawaii Telescope were performed with care and respect from the summit of Maunakea which is a significant cultural and historic site. Based on observations obtained with SPIRou, an international project led by Institut de Recherche en Astrophysique et Planétologie, Toulouse, France.
This work made use of \textsc{tpfplotter} by J. Lillo-Box (publicly available in \url{www.github.com/jlillo/tpfplotter}), which also made use of the python packages \textsc{astropy}, \textsc{lightkurve}, \textsc{matplotlib} and \textsc{numpy}.
This work makes use of observations from the LCOGT network. Part of the LCOGT telescope time was granted by NOIRLab through the Mid-Scale Innovations Program (MSIP). MSIP is funded by NSF.
This research has made use of the Exoplanet Follow-up Observation Program (ExoFOP; DOI: 10.26134/ExoFOP5) website, which is operated by the California Institute of Technology, under contract with the National Aeronautics and Space Administration under the Exoplanet Exploration Program.
Funding for the TESS mission is provided by NASA's Science Mission Directorate. KAC acknowledges support from the TESS mission via subaward s3449 from MIT.
This paper made use of data collected by the TESS mission and are publicly available from the Mikulski Archive for Space Telescopes (MAST) operated by the Space Telescope Science Institute (STScI). 
We acknowledge the use of public TESS data from pipelines at the TESS Science Office and at the TESS Science Processing Operations Center. 
Resources supporting this work were provided by the NASA High-End Computing (HEC) Program through the NASA Advanced Supercomputing (NAS) Division at Ames Research Center for the production of the SPOC data products.
This paper makes use of observations made with the MuSCAT2 instrument, developed by the Astrobiology Center, at TCS operated on the island of Tenerife by the IAC in the Spanish Observatorio del Teide.
The ULiege’s contribution to SPECULOOS has received funding from the European Research Council under the European Union’s Seventh Framework Programme (FP/2007-2013; grant Agreement no. 336480/SPECULOOS), from the Balzan Prize and Francqui Foundations, from the Belgian Scientific Research Foundation (F.R.S.-FNRS; grant no. T.0109.20), from the University of Liège, and from the ARC grant for Concerted Research Actions financed by the Wallonia-Brussels Federation. The Cambridge contribution is supported by a grant from the Simons Foundation (PI: Queloz, grant number 327127). The Birmingham contribution research is in part funded by the European Union’s Horizon 2020 research and innovation programme (grant’s agreement no. 803193/BEBOP), from the MERAC foundation, and from the Science and Technology Facilities Council (STFC; grant no. ST/S00193X/1, and ST/W000385/1). J.d.W. and MIT gratefully acknowledge financial support from the Heising-Simons Foundation, Dr. and Mrs. Colin Masson and Dr. Peter A. Gilman for Artemis, the first telescope of the SPECULOOS network situated in Tenerife, Spain. The University of Bern contribution is supported by a grant from the Swiss State Secretariat for Education, Research and Innovation (contract number MB22.00046, PI: Demory) as well as the Swiss National Science Foundation (grant IZSTZ0\_216537). The SPECULOOS North consortium would like to thank IAC telescope operators (Técnico de Operaciones Telescópicas), General and Instrumental maintenance teams for their support on site.
TRAPPIST-South is funded by the Belgian National Fund for Scientific Research (FNRS) under the grant PDR T.0120.21.
MG is FNRS Research Director. 
EJ is FNRS Senior Research Associate.
This publication benefits from the support of the French Community of Belgium in the context of the FRIA Doctoral Grant awarded to M.T.
The postdoctoral fellowship of KB is funded by F.R.S.-FNRS grant T.0109.20 and by the Francqui Foundation.

\section{Author Contributions}
E.M.B. lead one of the initial discovery efforts of the planet and was a key member in the ESPRESSO programme which measured the planet's mass. E.M.B. developed much of the text and figures and coordinated all the contributions. A.J. is the Principal Investigator of the ESPRESSO programme which measured the planet's mass. J.D.H. performed the joint analysis of all the data. A.J., J.D.H., D.B., V.V.E. all provided significant input into the text and figures. M.H. performed the spectral analysis of the ESPRESSO spectra. I.V.C. and B.V.R. obtained the Magellan/FIRE spectra and performed the relevant spectral analysis. K.B. performed the data reduction for the SPECULOOS and TRAPPIST observations and provided the text summarising the observations. J.C. performed the PandExo transmission spectra simulations. A.T. provided the text on the atmospheric characterisation potential of the planet. F.J.P. obtained the OSN observations and performed the data reduction for these observations and performed the analysis assessing the presence of and detection limits for additional planets. D.P.T. performed the interior structure analysis. K.N. performed an independent discovery of the planet.
J.D.H., D.B., E.S., G.\'A.B., R.B., M.H. all contributed to the ESPRESSO programme under which the ESPRESSO data were obtained. 
J.M.A. and X.B. obtained the ExTrA photometry. 
K.A.C. co-ordinated the TFOP SG1 photometric follow-up campaign. K.A.C., K.I.C., G.S. contributed to the LCOGT observations through scheduling, observing and data reduction.
T.G., C.Ca., L.A. contributed to the SPIRou observations through planning, scheduling, and data reduction.
S.B.H., C.Cl., R.G., C.L. obtained the Gemini speckle imaging.
G.F.-R., A.F., I.F., K.I., F.M., N.N., E.P., H.P. contributed to the MuSCAT2 photometric facility and observations.
K.B., A.B., B.-O.D., G.D., E.D., M.G., M.J.H., E.J., F.J.P., D.Q., B.V.R., D.S., A.T., M.T., J.d.W., S.Z.-F. contributed to the SPECULOOS and TRAPPIST facilities and observations through planning, management, scheduling, data collection, and data reduction.
P.J.A., V.C., F.J.P. contributed to the OSN facility and observations through management, planning, data collection and reduction.
D.A.C., D.C., T.D., J.M.J., A.M.L., S.S., R.V., J.N.W. provided essential contributions to the TESS mission, which provided the discovery data for the planet.
All authors read the manuscript and provided comments on the content and wording.

\section{Competing interests}
The authors declare no competing interests.








\bibliography{tic67512645}
\bibliographystyle{sn-basic}




\end{document}